\documentclass[%
 reprint,
 amsmath,amssymb,
 aps,
floatfix,
]{revtex4-1}
\usepackage[english]{babel}
\usepackage{blindtext}
\usepackage{graphicx}
\usepackage{dcolumn}
\usepackage{bm}

\newcommand{\mo}{{-1}}

\begin{document}

\preprint{APS/123-QED}

\title{Market fragmentation and market consolidation: multiple steady states in systems of adaptive traders choosing where to trade}

\author{Aleksandra Alori\'c}
 \email{aleksandra.aloric@gmail.com}
\affiliation{
Scientific Computing Laboratory, 
Center for the Study of Complex Systems\\
Institute of Physics Belgrade, 
University of Belgrade, Pregrevica 118, 11080 Belgrade, Serbia
}%
\author{Peter Sollich}
\affiliation{%
Institut fur Theoretische Physik, Georg-August-Universit\"at G\"ottingen\\
Friedrich-Hund-Platz 1,
D-37077 G\"ottingen, Germany\\
Disordered Systems Group, Department of Mathematics, King's College London, Strand, London WC2R 2LS, UK
}%

\begin{abstract}
Technological progress is leading to proliferation and diversification of trading venues, thus increasing the relevance of the long-standing question of market fragmentation versus consolidation. To address this issue quantitatively, we analyse systems of adaptive traders that choose where to trade based on their previous experience. We demonstrate that only based on aggregate parameters about trading venues, such as the demand to supply ratio, we can assess whether a population of traders will prefer fragmentation or specialization towards a single venue. We investigate what conditions lead to market fragmentation for populations with a long memory and analyse the stability and other properties of both fragmented and consolidated steady states. Finally we investigate the dynamics of populations with finite memory; when this memory is long the true long-time steady states are consolidated but fragmented states are strongly metastable, dominating the behaviour out to long times.

\end{abstract}

\maketitle

\section{\label{sec:Intro}Introduction}

Whether a consolidated or a fragmented market is more beneficial to a population of traders is a long-standing debate~\cite{mendelson1987consolidation,Chowdhry1991,madhavan1995,biais2005market,Bennett2006,degryse2015}. In a consolidated or concentrated market, the majority of trades occurs in one (or a few) as opposed to numerous trading venues. With technological advances we have seen a proliferation of trading venues such as online marketplaces. Even more recently alternative or dark trading venues have appeared, e.g.\ \textit{dark pools}. These are popular not least for their lack of transparency, which makes them interesting for trading large quantities of shares without strongly influencing the price, see e.g.~\cite{darkpool,degryse2015}. 

The emergence of collective behaviour in systems of autonomous agents is a research topic that has seen widespread interest among physicists in the last couple of decades. The main reason for this is the recognition that statistical physics techniques, which contributed to the understanding of macroscopic phenomena arising in large systems of interacting microscopic entities, can be applied to a range of biological, economic and social systems. A large body of work exists in the physics literature on collective effects in socio-economic systems~\cite{castellano2009,bouchaud2019econophysics}, e.g.\ mass movement of people~\cite{helbing1995,moussaid2011}, herd behaviour of traders~\cite{tedeschi2012herding}, voting patterns~\cite{galam1982,fernandez2014}  etc. One of the most prominent examples is the minority game, which continues to attract interest due to its simplicity and its ability to  reproduce at least ``stylized'' facts about financial markets \cite{challet2001, challet2003}; extensions of the model also predict interesting grouping phenomena when multiple assets are available to agents \cite{Huang2012}. In a similar vein, in this paper, we investigate whether fragmentation and consolidation can arise solely as a consequence of interactions at the level of the agents, combined with individual adaptation. 

Some studies of stylized models of market competitions already exist, often pointing out the emergence of monopolies whereby the majority of trades occurs in one trading venue. Pagano~\cite{pagano1989trading} argues that when markets are identical (in terms of their transaction costs), risk averse traders will concentrate in a single market. On the other hand, when there is asymmetry, fragmentation might arise with traders being clustered based on the sizes of their desired transactions. Chowdhry and Nanda~\cite{Chowdhry1991} reach the same conclusion in a system with asymmetrically informed traders and a general number of markets.

Ellison \textit{et al.}~\cite{ellison2004} and Shi \textit{et al.}~\cite{shi2013} also study competition among markets and the conditions under which such competition can lead either to monopolies or to coexistence of multiple markets. The authors name two significant effects in a competition of double auction trading venues. One of them is the positive size effect, i.e.\ agents prefer to trade in a market where there are already many traders of the opposite type. As an example, sellers like trading at markets where there are many buyers as this gives them a wider choice of offers. The authors of Refs.~\cite{ellison2004,shi2013} also suggest the existence of a negative size effect in a double auction market: agents will prefer being in the minority group of traders more often, with e.g.\ buyers benefiting from trading at a market where there are not many other buyers, see e.g.~\cite{caillaud2003chicken}). Ellison \textit{et al.}~\cite{ellison2004} point out that such negative size effects can enable the coexistence of many markets. On the other hand Shi \textit{et al.}~\cite{shi2013} investigate which of the two effects is stronger and find that due to more substantial positive effects, a monopoly will be the favoured end-state in many situations. The authors of~\cite{shi2013} argue that market coexistence remains a possibility when there is strong market differentiation, especially for markets that have different pricing policies: one market might charge a fixed participation fee while another might take a profit fee. A common feature of the studies mentioned above is that some form of Nash equilibrium analysis was used, assuming perfect information about the activity of all traders and maximisation of an underlying utility function for the agents. 

The increased proportion of trades that take place in dark trading venues -- 15\% of the US share market volume was traded in dark pools in 2013~\cite{shorter2014dark} -- suggests increased fragmentation at least between traditional and dark trading. Calling for more research on reasons behind market fragmentation, Gomber \textit{et al.}~\cite{gomber2017competition} suggest the heterogeneity of traders and their needs as one of the main drivers of market fragmentation.
However, studies of similar effects such as the emergence of market loyalty in fish markets~\cite{Kirman2000}, herding~\cite{tedeschi2012herding} and grouping of agents in multi-resource minority games~\cite{Huang2012} show that fragmentation-like phenomena may also be emergent. Nonetheless, existing models for such emergent fragmentation often assume a considerable amount of structure, e.g.\ in the connectivity among agents, the information available about the actions of other players, the rules of interaction via the market mechanism, the asymmetry between buyers and sellers, etc. In contrast, we study here a model in which initially homogeneous agents adapt only to their private information, and show that even in such a system both market fragmentation and consolidation can occur depending on global system parameters.

Based on observations from the CAT tournament~\cite{cai2009overview}, where the spontaneous emergence of long-lived market loyalties was seen in complicated systems of adaptive markets and traders, we hypothesize that the reason for fragmentation may not lie in the intricacies of different market mechanisms or trading strategies. Instead, we conjecture that fragmentation is a collective phenomenon arising as a consequence of the continuous adaptation of the individual agents to an evolving system. To test this hypothesis, we developed a stylised model of double auctions and adaptive traders~\citep{Aloric2015,aloric2016} that does indeed predict emergent fragmentation under minimal assumptions on the complexity of market and trading mechanisms. The model also shows market consolidation under some circumstances. Our focus in this study is to pin down under what conditions fragmentation and consolidation occur, and what relative benefits they bring for the tradners. As we will see the behaviour of the model is remarkably rich in spite of its simplicity, with multiple steady states coexisting in the limit of long agent memory. For finite memory length, this can lead to the existence of long-lived metastable states that dominate before the true steady state is reached eventually.

We start with a short description of the model \cite{aloric2016} in section~\ref{sec:model} and then proceed to the large memory limit analysis of small systems with $N=2$ and 4 agents in section~\ref{sec:finiteN}. These can be thought of as two- and four-player \textit{games}. They are convenient as we can easily track each trader's adaptation. At the same time they already reveal qualitative phenomena related to those we find later in large systems, in particular coordination at the same market (for $N=2$) and onset of fragmentation via pairwise coordination (for $N=4$). Moving on to the large population limit ($N\to\infty$), we then first analyse a \textit{population with homogeneous buying preferences}~\ref{sec:singlepopulation}. We develop the relevant mathematical framework and techniques of analysis here and then generalise the results to systems with separate buyer and seller agent types. Finally we study the system dynamics in some detail to go beyond the steady state analysis in section~\ref{sec:2populations}.

Overall we follow a typical statistical physics philosophy in using a model that reduces the underlying market choice dynamics to its key ingredients, allowing us to obtain detailed insights into the origins of the resulting collective behaviour. The analysis also relies significantly on statistical physics concepts and methods: we focus mostly on the thermodynamic limit of large agent populations, where we exploit the fact that the behaviour of $N$ interacting agents for $N\to\infty$ can be captured by the dynamics of a single agent subject to self-consistently determined population-level order parameters. The main outcome from this physical point of view is the emergence of multiple non-trivial steady states in the large interacting non-equilibrium systems that we study.

\section{\label{sec:model}Model}
Here we summarize basic assumptions and properties of the model introduced in \cite{Aloric2015,aloric2016}, which is the foundation for the analysis in this paper.

\textbf{\textit{Learning.}} In the model, agents choose among the available markets once in every trading period and submit their order to the chosen market. A key assumption is that agents base their decision of where to trade on their previous experience at the different markets. Agents rely on the following reinforcement rule, which is based on the experience-weighted attraction rule~\cite{EWACamerer1999,EWAHo2007} but neglects knowledge about the other markets (via so-called fictitious payoffs):
\begin{small}
\begin{equation}
  \label{eq:3M:EWAupdate}
  A^i_m(n+1) =\left\{
    \begin{array}{cc}
      (1 - r) A^i_m(n) + r S^i_m(n), & \text{$m$ chosen in round $n$}\\
      (1 - r) A^i_m(n), & \text{otherwise}
    \end{array}\right.
\end{equation} 
\end{small}
Here $A^i_{m}(n+1)$ is agent $i$'s attraction to market $m$ at trading period $n+1$ given his/her score or return, $S^i_m(n)$, obtained in the previous trading period (see below) and the previous attraction $A^i_m(n)$. To understand the role of $r$ one can write down the resulting general expression for the attraction at trading round $n$:
\begin{small}
\begin{align*}
 A^i_m(n)&=\sum_{j=0}^{n-1} r(1-r)^{n-j}\delta_{m^i(j),m}S^i_m(j)\\
 &+(1-r)^{n}A_m^i(0)
\end{align*} 
\end{small}
where the Kronecker $\delta$ restricts updates to rounds where the agent's chosen market $m^i(j)$ is the one ($m$) being considered. The factor $r(1-r)^{n-j}$ in this expression is a weight that decays exponentially into the past, becoming small once $n-j$ is of order $1/r$. Thus each agent effectively averages scores over a sliding window into the past of length $\approx 1/r$, so that $1/r$ can be thought of as setting the length of the agents' memory.

To choose a market at each trading round, an agent translates the learned  attractions into probabilities of choosing each markets, using the \emph{multinomial logit} or softmax function:
\begin{equation}
  \label{eq:3M:MultinomialLogit}
  P_m^i(n)  = \frac{\exp(\beta A_m^i(n))}{\sum_{m'} \exp(\beta A_{m'}^i(n))}
\end{equation}
This aspect of the model is also in line with the experience-weighted attraction literature~\cite{EWACamerer1999,EWAHo2007}; $\beta$ is the \textit{intensity of choice} and regulates how strongly the agents bias their preferences towards actions with high attractions. For $\beta\rightarrow\infty$ the agents choose the option with the highest attraction, while for $\beta\rightarrow 0$ they choose randomly and with equal probabilities among all options.

We study agents whose choice of the \textit{type} of trading order (to buy or to sell) is not adaptive but rather set by a fixed buying preferences $p^i_{\mathcal B}$. This assumption simplifies the analysis while still allowing both consolidation and fragmentation behaviour as shown previously~\cite{aloric2016}.

\textbf{\textit{Trading strategies.}} Agents do not have sophisticated trading strategies in our model and are essentially zero-intelligence traders~\cite{Gode1993,duffy2006,ladley2012zero}. Their orders to buy (bid) or sell (ask) a single unit of the underlying good at a certain price are independent of previous returns or other information. We assume specifically that bids, $b$, and asks, $a$, are normally distributed as $a \sim \mathcal{N}(\mu_a,\sigma_a^2)$ and $b \sim \mathcal{N}(\mu_b,\sigma_b^2)$, where we fix $\mu_b>\mu_a$ as in~\cite{aloric2016}. 
After each round of trading each agent receives a score, reflecting their payoff in the trade. This depends on the global trading price set by a chosen market $m$ as well as the order the agent has submitted. The scores of agents who do trade are assigned as in previous studies~\cite{Gode1993,Anufriev2013}: buyers value paying less than they offered ($b$), and so their score is $S=b-\pi$. Sellers value trading for more than their ask ($a$), and so  $S=\pi-a$ is a reasonable model for their payoff; in both cases $\pi$ is the trading price. 

\textit{\textbf{Market mechanism.}} In the spirit of keeping the model as simple as possible we consider double auction markets in discrete time, counted as before in trading rounds. In every round the global trading price is set by the market: once all orders have arrived, these are used to determine the average bid $\langle b\rangle$ and average ask $\langle a\rangle$ and set the price:
\begin{equation} 
\pi=\langle a\rangle+\theta(\langle b\rangle-\langle a\rangle)
\label{price}
\end{equation} 
where $\theta$ fixes the price closer to the average bid ($\theta>0.5$) or the average ask ($\theta<0.5$), as in~\cite{Aloric2015}. This parameter thus represents the bias of the market towards buyers or sellers. Once the trading price has been set, all bids below this price, and all asks above it, are marked as invalid orders as they cannot be executed at the current trading price. The remaining orders are executed by randomly pairing buyers and sellers. 
Excess buyers or sellers, i.e.\ those that cannot be paired, receive zero score, as do the agents who submitted invalid orders.

Note that traders are not informed about the market biases, nor the market mechanism in general. The only information they have at their disposal in order to adapt their market preferences is their personal score.

\section{\label{sec:finiteN}Finite N}

\subsection{\label{sec:2players}Two traders: coordination}

To understand collective effects in trading systems, we first build up some intuition by looking at a very simple model with only one buyer and one seller. The traders have a choice between two markets with different biases. As the system consists of only two agents and two markets, fragmentation (or segregation as introduced previously~\cite{aloric2016}), in which a population will split into distinctive groups favouring one option, is not feasible. However, we can investigate if long lasting loyalty to a single market emerges, which can signal market consolidation.

To make trading possible the two agents effectively need to coordinate, i.e.\ to submit orders to the same market. This can lead to one of the agents earning less than they could have done at the other market. One question of interest concerns the conditions under which the agents prefer random decisions of who will be a \textit{winner} or \textit{loser} in this manner, as opposed to settling in these roles over longer periods of time. Thus we will focus on the existence of \textit{coordination of traders} and investigate for which parameter settings agents develop strong preferences for the same market. Intriguingly, this two player analysis ends up being largely similar to the work by Hanaki~\textit{et al.}~\cite{Hanaki2011} where a two-agent case was likewise studied as a first step to understanding collective effects. (In~\cite{Hanaki2011} these concerned specialization behaviour of agents searching for parking spots.)

For the $N=2$ analysis it is convenient to label the two players as $i=\pm 1$ and similarly for the two markets. We use the following specific parameter settings:
\begin{itemize}
\item Of the two players, player $i=1$ always buys while player $i=\mo$ always sells ($p^1_{\mathcal B}=1$, $p^\mo_{\mathcal B}=0$). 
\item Bids/asks are deterministic, i.e.\ $b\sim \mathcal{N}(\mu_b,0)$, $a\sim \mathcal{N}(\mu_a,0)$, with their difference being fixed to $\mu_b-\mu_a=1$.
\item The trading price at each market is set as defined in~\cite{Aloric2015}, $\pi_m=\langle a\rangle+\theta_m(\langle b\rangle-\langle a\rangle)$.
\item We assume that the market biases are \textit{symmetric}: $(\theta_1,\theta_\mo)=(\theta,1-\theta)$ where $\theta\in[0,0.5]$.
\end{itemize}
The simplification over our previous work~\cite{Aloric2015,aloric2016} of making bids and asks  deterministic allows us to focus solely on the coordination of the market choices and does not change the behaviour of the system qualitatively. 
The determinstic order prices then also make the trading prices deterministic: $\pi_m=\mu_a+\theta_m(\mu_b-\mu_a)=\mu_a+\theta_m$.

We can summarize the attraction update rule ~\ref{eq:3M:EWAupdate} as
\begin{align*}
A^i_m(n+1)=(1-r)A_m^i(n) + rS_m^i(n)\ ,
\end{align*}
with the convention that $S_m^i(n)=0$ if market $m$ was not chosen by agent $i$ in round $n$. This generalized score is fully determined by the market choice of the opposite player:
\begin{align}
S_m^i(n)&=
\delta_{m^{i}(n),m}
\delta_{m^{-i}(n),m}
\Sigma_m^i\ ,
\label{eq:3M:2pscore}
\end{align}
where $m^{(-)i}(n)$ denotes the market of choice of the (co)player $(-)i$ during trade $n$ and
\begin{align}
\Sigma_m^i&=\begin{cases}
\mu_b-\pi_m=1-\theta_m, & i=1\\
\pi_m-\mu_a=\theta_m, & i=\mo 
\end{cases}
\label{eq:3M:2playersigma}
\end{align}
encodes the relevant nonzero score values that depend on the type of market and agent. 
The logit assignment (\ref{eq:3M:MultinomialLogit}) by which agents choose a market $m$ simplifies for $N=2$  to
\begin{align*}
P^i_m(n)
=\frac{1}{1+\exp\left(-\beta m\Delta^i(n)\right)} = \sigma_\beta(m\Delta^i(n))\ ,
\end{align*}
where $\sigma_\beta(z)=[1+\exp(-\beta z)]^{-1}$ is the logistic sigmoid. The choice probabilities do not depend on the attractions to the two markets individually but only on their difference $\Delta^i=A^i_1-A^i_{\mo}$. The latter is updated as
\begin{align*}
\Delta^i(n+1)&=A^i_1(n+1)-A^i_{\mo}(n+1)\\
&=rS^i_1(n)+(1-r)A^i_1(n)\\
&-\left[rS^i_{\mo}(n)+(1-r)A^i_{\mo}(n)\right]\ .
\end{align*}
The stochastic variable $\Delta^i(n+1)$ thus depends on the choices the agents make in trading round $n$, $m^i(n)$ and $m^{-i}(n)$, which are drawn from distributions that depends on $\Delta^i(n)$ and $\Delta^{-i}(n)$. This situation simplifies in the long memory limit $r\rightarrow 0$, where the attraction differences change sufficiently slowly to average out stochastic fluctuations. One can then effectively replace $\delta_{m^i(n),1}$ by its expected value $\sigma_\beta\left(\Delta^i(n)\right)$ (and similarly for $-i$ and other market choices) in the score (\ref{eq:3M:2pscore}). This gives
\begin{align*}
&\Delta^i(n+1)=r\Big[\sigma_\beta\left(\Delta^i(n)\right)\sigma_\beta\left(\Delta^{-i}(n)\right)\Sigma_1^i\\
&-\sigma_\beta\left(-\Delta^i(n)\right)\sigma_\beta\left(-\Delta^{-i}(n)\right)\Sigma_{\mo}^i\Big]
+(1-r)\Delta^i(n)\ ,
\end{align*}
which can be further simplified into:
\begin{align*}
&\frac{\Delta^i(n+1)-\Delta^i(n)}{r}\\
&=-\Delta^i(n)+\Big[\sigma_\beta\left(\Delta^i(n)\right)\sigma_\beta\left(\Delta^{-i}(n)\right)\Sigma_1^i\\
&-\sigma_\beta\left(-\Delta^i(n)\right)\sigma_\beta\left(-\Delta^{-i}(n)\right)\Sigma_{\mo}^i\Big]\ .
\end{align*}
The finite difference on the l.h.s.\ becomes a derivative in the limit of small $r$ if we switch to the rescaled time $t=nr$, for which a unit time interval corresponds to $1/r$ trading periods:
\begin{align*}
\partial_t\Delta^i(t)=&-\Delta^i(t)+\Big[\sigma_\beta\left(\Delta^i(t)\right)\sigma_\beta\left(\Delta^{-i}(t)\right)\Sigma_1^i\\
&-\sigma_\beta\left(-\Delta^i(t)\right)\sigma_\beta\left(-\Delta^{-i}(t)\right)\Sigma_{\mo}^i\Big]\ .
\end{align*}
A convenient change in variables that simplifies this pair of differential equations is  $\Delta^1(t)=\xi(t)+\rho(t)$ and $\Delta^{\mo}(t)=\xi(t)-\rho(t)$, which after some algebra and exploiting the market symmetry gives
\begin{align}
\label{eq:3M:2pdynamicaleq}
\partial_t \xi(t)&=-\xi(t)+\frac{1}{2}\frac{\sinh\left(\beta \xi(t)\right)}{\cosh\left(\beta\xi(t)\right)+\cosh\left(\beta\rho(t)\right)}\ ,\\
\partial_t \rho(t)&=-\rho(t)+\frac{1-2\theta}{2}\frac{\cosh\left(\beta\xi(t)\right)}{\cosh\left(\beta\xi(t)\right)+\cosh\left(\beta\rho(t)\right)}\ .
\nonumber
\end{align}
Note that $\xi=(\Delta^{1}+\Delta^{\mo})/2$ describes the average of the attraction differences of the two agents, while $\rho=(\Delta^{1}-\Delta^{\mo})/2$ captures the deviation between them.

To understand the dynamics we first consider its fixed points, which need to satisfy
\begin{align}
\label{eq:3M:2pfixedpoint}
\xi^*&=\frac{1}{2}\frac{\sinh\left(\beta\xi^*\right)}{\cosh\left(\beta\xi^*\right)+\cosh\left(\beta\rho^*\right)}\ ,\\
\rho^*&=\frac{1-2\theta}{2}\frac{\cosh\left(\beta\xi^*\right)}{\cosh\left(\beta\xi^*\right)+\cosh\left(\beta\rho^*\right)}\ .\nonumber
\end{align}
The first of these equations is always satisfied if $\xi^*=0$, and in that case the equation for $\rho^*$ has a unique solution whose sign depends on the sign of $1-2\theta$. When market 1 is favourable towards buyers ($\theta<0.5$), $\rho^*$ will be positive. As $\Delta^{\pm 1}=\pm \rho^*$, this can be interpreted as a state where buyers and sellers learn which market is good for them, and thus have preferences for opposite markets. ($\Delta^1$ is positive meaning that player 1, the buyer, prefers market 1, which is good for buyers.) As we shall see shortly this solution is only stable for low intensities of choice where the agents' market choice dynamics remains largely random. The intuition for the appearance of an instability with increasing $\beta$ is that, if agents were to follow through fully on their attractions towards opposite markets, they would never get to trade.

The stability of the solution $(\xi^*=0,\rho^*)$ can be studied by linearizing the dynamical equations~(\ref{eq:3M:2pdynamicaleq}), resulting in the stability criterion
\begin{align*}
\frac{\beta}{2}\frac{1}{1+\cosh(\beta\rho^*)}\leq1\ .
\end{align*}
Expressed in the original variables $\Delta^i$, the solution with $\Delta^{1*}+\Delta^{\mo*}=0$ is stable as long as
\begin{equation}
\frac{\beta}{2}\frac{1}{1+\cosh(\beta(\Delta^{1*}-\Delta^{\mo*})/2)}\leq1\ .
\label{stability}
\end{equation}
This stability condition is exactly the same as in Ref.~\cite{Hanaki2011} because the learning dynamics we follow is essentially the same and differs only in the details of the deterministic returns.

\begin{figure}[h!]
        \centering
\includegraphics[width=0.48\textwidth]{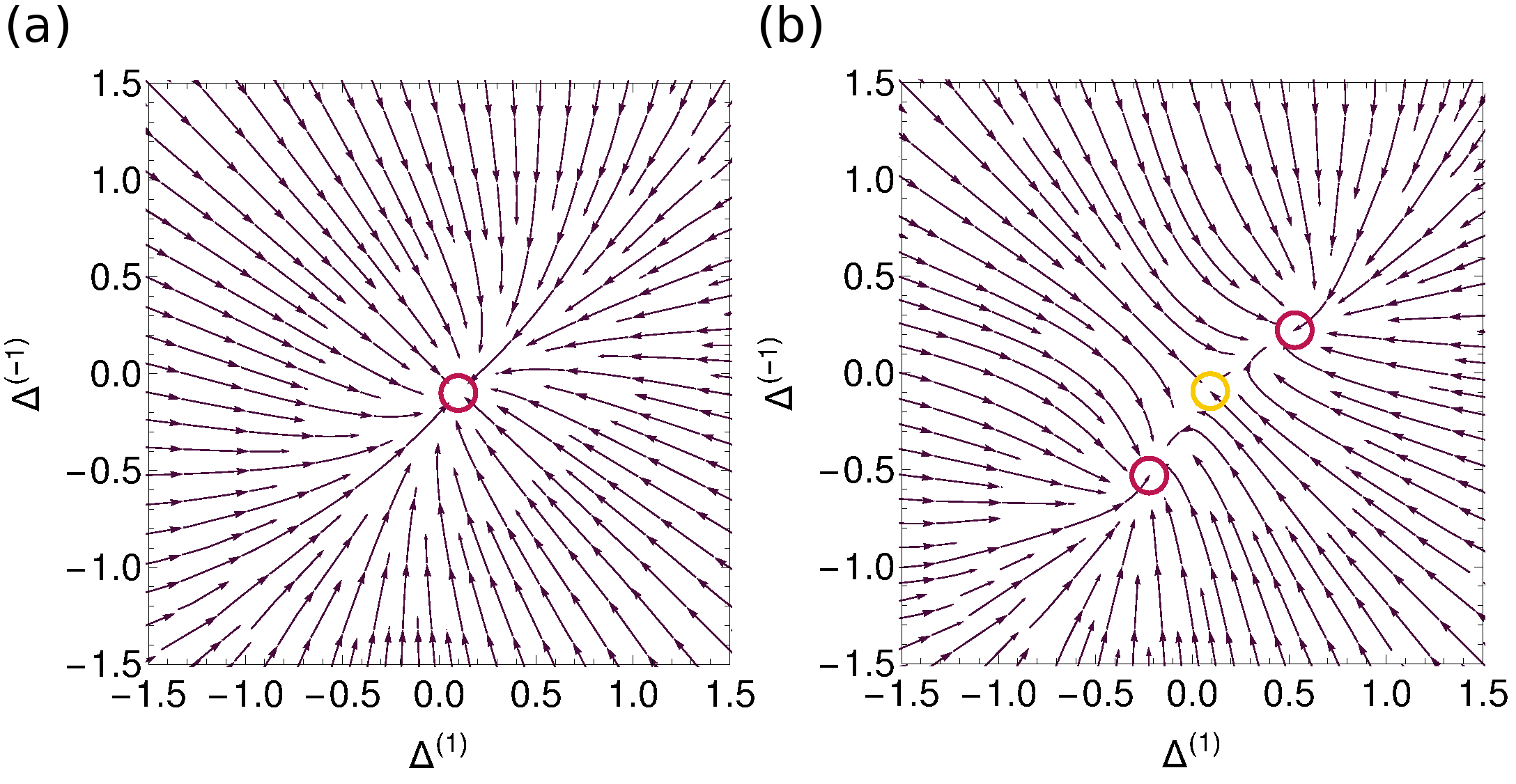}
\caption[2 player-dynamics: flow diagrams]{Two-trader dynamics: flow diagrams (\ref{eq:3M:2pdynamicaleq}) for (a) the intensity of choice $\beta=2$, with a unique fixed point where agents decide largely randomly, and (b) $\beta=6$, with two new fixed points indicating where coordinated states appear. For the market bias used, $\theta=0.3$, the critical intensity of choice where coordinated states emerge is $\beta_c=4.16$.}
\label{flowdiags}
\end{figure}

We illustrate in Figure~\ref{flowdiags} that for low intensities of choice, where the stability criterion (\ref{stability}) is satisfied, the fixed point discussed so far is the only one. At higher $\beta$ the criterion is violated and two new stable fixed points appear.
Here the agents' attraction differences are of the {\em same} sign, i.e.\ they prefer going to the same market. This happens even though market $1$ favours buyers while market $\mo$ favours sellers. 

At first sight it may seem puzzling that for high intensity of choice, one of the agents decides to settle for less in persistently choosing the market where (s)he will be awarded lower scores. However, this pattern of behaviour in fact maximizes the number of trades that take place. In the low-$\beta$ regime, all four pairings of market choices are equally probable: $(m^1,m^\mo)\in\{(1,1),(1,\mo),(\mo,1),(\mo,\mo)\}$ but only the first and the last enable trading. On the other hand, in the high $\beta$ regime, when both agents persistently choose the same market, they always get to trade, although one of the traders always receives a lower return. For the market parameters used in Fig.~\ref{flowdiags}, $(\theta_1,\theta_\mo)=(0.3,0.7)$, the agent who settles for a lower score then receives a score of $0.3$, while the other one obtains 0.7. This has to be compared to the average payoff at low $\beta$, which by averaging over the four market choice pairings is seen to be $\frac{1}{4}(\theta_1+\theta_\mo)=\frac{1}{4}$. Hence both agents clearly earn more in the coordinated regime than by choosing randomly.

\begin{figure}[h!]
\center
\includegraphics[width=0.48\textwidth]{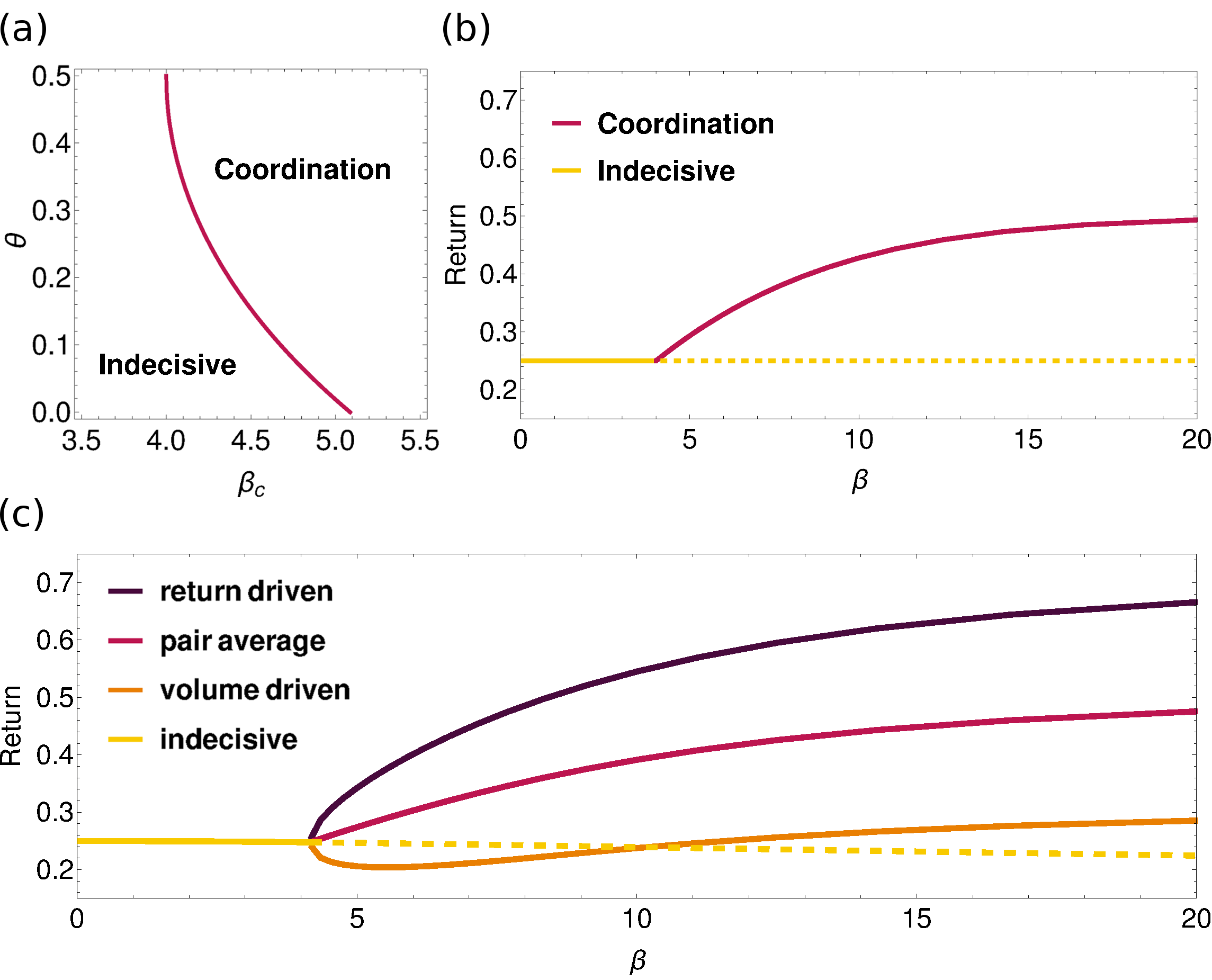}
        \caption[2 players: $(\theta,\beta)$-phase diagram and returns]{Two traders: the $(\theta,\beta)$ phase diagram and returns. (a) Coordination and indecisiveness regions for different intensity of choice and market biases ($\beta$ and $\theta$). (b) Returns for different $\beta$ in a system with two fair markets $\theta=0.5$. (c): Returns for different $\beta$ for market bias $\theta=0.3$. At the critical $\beta_c=4.16$, the average return of the two agents in the coordinated state is higher than it would be in the continuation of the low $\beta$ fixed point (yellow dashed line), but one of the agents needs to settle for less.}
\label{toymodelphasediag}
\end{figure}

We can find the domain of parameters $\theta$ and $\beta$ where the agents will coordinate (Figure~\ref{toymodelphasediag} (a)) by starting from the regime of agents choosing largely randomly and tracking where the stability condition (\ref{stability}) is first violated as $\beta$ is increased.
As in the case of large populations~\cite{aloric2016}, we observe that $\beta_c$ increases with increased market difference or bias. The symmetry breaking between markets that coordination requires is therefore {\em not} driven by the difference between the markets. In fact the coordination threshold is lowest for a system with two identical markets ($\theta=0.5$). One can rationalize this by saying that the agents are happiest to coordinate at one of the markets in this limit as neither needs to settle for less. We show average returns for this setup -- a pair of traders choosing between two unbiased markets -- as a function of $\beta$ in Figure~\ref{toymodelphasediag} (b). One observes the expected average score of $1/4$ for low $\beta$; 
as $\beta$ is increased, the agents effectively realize that coordination at a single market enables more trades and consequently higher average returns. 

In the panel (c) of Figure~\ref{toymodelphasediag} we show analogous results for the case of two biased markets ($\theta=0.3$). 
We plot the individual agents' payoffs and their average in the state where they coordinate at one market, and compare this to the payoff in the largely random low-$\beta$ state. As a reference we also plot the continuation of the latter to larger $\beta$, where it is unstable. It is notable that returns {\em decrease} with $\beta$ on this branch: the more the agents act on their preference for opposite markets, the less often they manage to meet at the same market. This results in more and more trading rounds where both receive a return of zero, dragging down average returns. 

By contrast, in the coordinated state the average return {\em increases} with $\beta$, i.e.\ as the agents make more and more definite choices. Interestingly, Figure~\ref{toymodelphasediag} (c) shows that this increase in the average return is accompanied by a growing difference between the returns of the individual agents. These payoff differences can occur in our model because agents are unaware of the opposite player's return, making decisions only on the basis of their own scores.
Borrowing terminology from the large system limit~\cite{aloric2016} we will call the agent with the higher return \textit{return driven} and the other \textit{volume driven}.
It is notable in Figure~\ref{toymodelphasediag} (c) that there is a range of $\beta$ where the volume driven agent receives an average return that is lower not only than that of the return oriented agent, but also than the hypothetical return both agents would achieve in the (unstable) uncoordinated state; this regime grows as the markets become more biased.

Intuitively, the return driven player develops a strong preference for the market where (s)he can earn more. The other agent will occasionally try the other market, but typically not get to trade there. As this results in a zero return, (s)he is better off persisting with the coordinated choice, which offers a low but at least nonzero return.

The two agent systems studied so far can be mapped to two player games: the \textit{symmetric pure coordination game} when the markets are unbiased, and the \textit{battle of the sexes} when markets are symmetrically biased. For these games it is known that the two coordinated states correspond to pure Nash equilibria, see e.g.~\cite{easley2010networks}. In the symmetric pure coordination game, both of these are envy free (i.e.\ both agents earn the same), but not so in the battle of the sexes -- this is consistent with the differences we saw between unbiased and biased markets, and the Nash equilibria correspond to the $\beta\to\infty$ limit of the coordinated states. There are also mixed Nash equilibria. These correspond to the continuation to $\beta\to\infty$ of our uncoordinated state for the symmetric pure coordination game, but not otherwise. A full correspondence to Nash equilibria could be obtained by modifying the learning rule so that the attractions to markets that were not chosen are kept unchanged. This can be interpreted as ``fictitious play'' and is discussed in more detail in~\cite{Nicole2018dynamical}.

The results described above can be generalized to a pair of traders who do not have strict buyer and seller roles but instead decide to buy with some probability. We assume symmetric preferences for buying, $p_\mathcal{B}^1=1-p_\mathcal{B}^\mo=p_\mathcal{B}$. 
For a trade to occur, agents now need to be at the same market {\em and} need to submit opposite (buy and sell) orders. As the buying preferences $p_\mathcal{B}^i$ are fixed, this only changes $\Sigma_m^i$ from (\ref{eq:3M:2playersigma}) to
\begin{align}
\Sigma_m^i=p_\mathcal{B}^i(1-p_\mathcal{B}^{-i})(1-\theta_m)+(1-p_\mathcal{B}^i)p_\mathcal{B}^{-i}\theta_m\ .
\label{deterministiscore}
\end{align}
To see this, note that agent $i$ receives a buyer payoff $1-\theta_m$ when (s)he assumes the role of a buyer (with prob $p_\mathcal{B}^i$) while the opposite player acts as seller; (s)he also receives a seller payoff in the opposite configuration. Repeating the calculation above one then finds that the fixed point conditions (\ref{eq:3M:2pfixedpoint}) both acquire a factor of $p_\mathcal{B}^2+(1-p_\mathcal{B})^2$ on the r.h.s., while the stability condition (\ref{stability}) is multiplied by the same factor.
\begin{figure}
\centering
\includegraphics[width=0.5\textwidth]{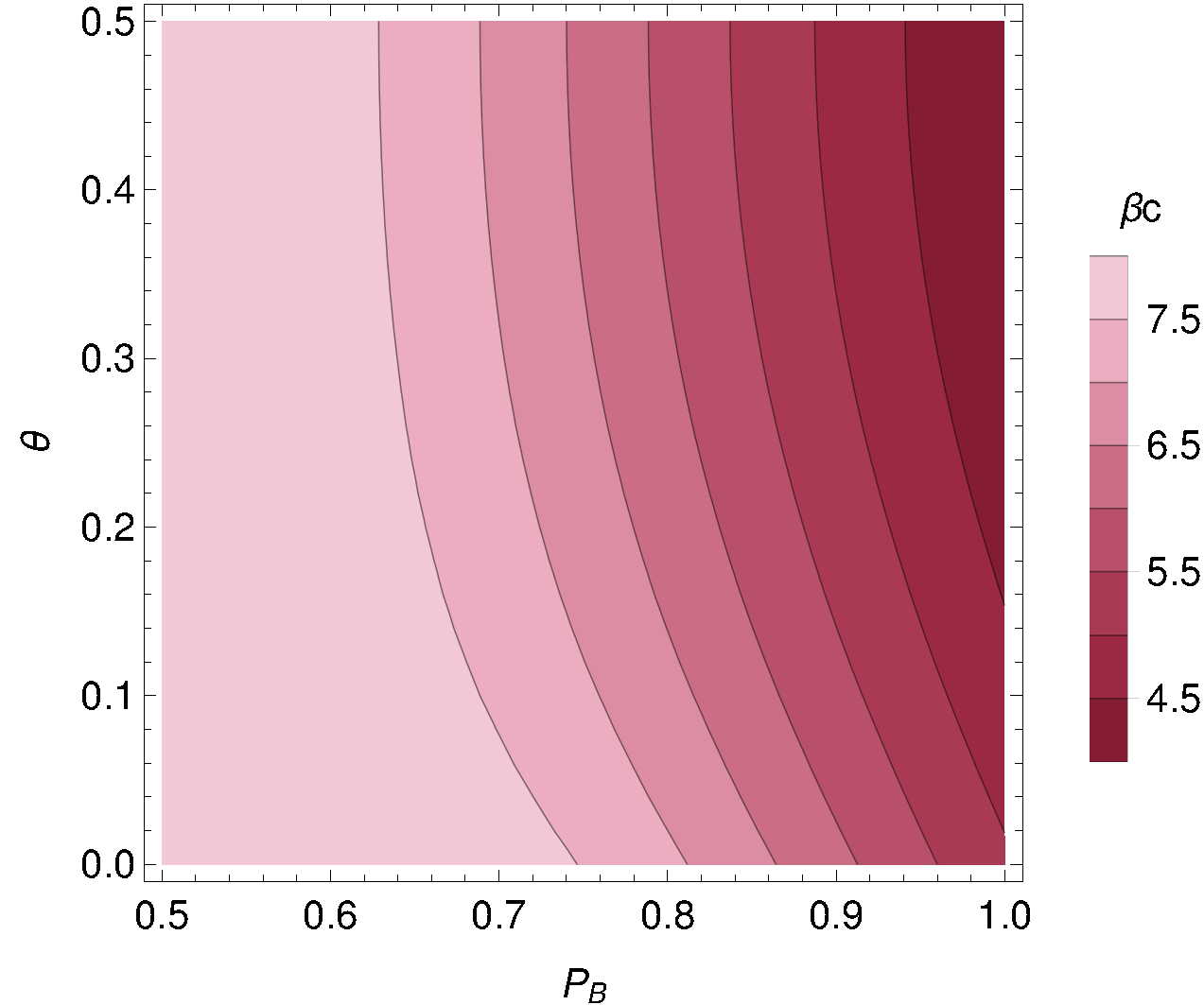}
\caption[Two players: coordination threshold as a function of $\theta$ and $p_\mathcal{B}$]{Two traders: coordination threshold as a function of $\theta$ and $p_\mathcal{B})$. Note that the threshold is finite for all system parameters and increases the more similar the agents become, i.e., as $p_\mathcal{B}$ decreases towards 0.5.}
\label{thpbdiagram}
\end{figure}
Figure \ref{thpbdiagram} shows contours of the resulting critical $\beta_c$ for coordination.
We note that the coordination threshold increases as $p_\mathcal{B}$ approaches 1/2: agents without strong buy/sell preferences need higher intensities of choice to benefit from the coordinated state. This makes sense because agents with $p_\mathcal{B}$ closer to $1/2$ derive a lower benefit from coordinating at a market: as they need to assume buyer and seller roles, trades at the same market happen only with some probability, specifically $p_\mathcal{B}^2+(1-p_\mathcal{B})^2$ in our setting, which approaches one half for $p_\mathcal{B}\to 1/2$.

\subsection{\label{sec:4players}Four traders: onset of fragmentation}

\begin{figure*}[t]
\centering
\includegraphics[width=\textwidth]{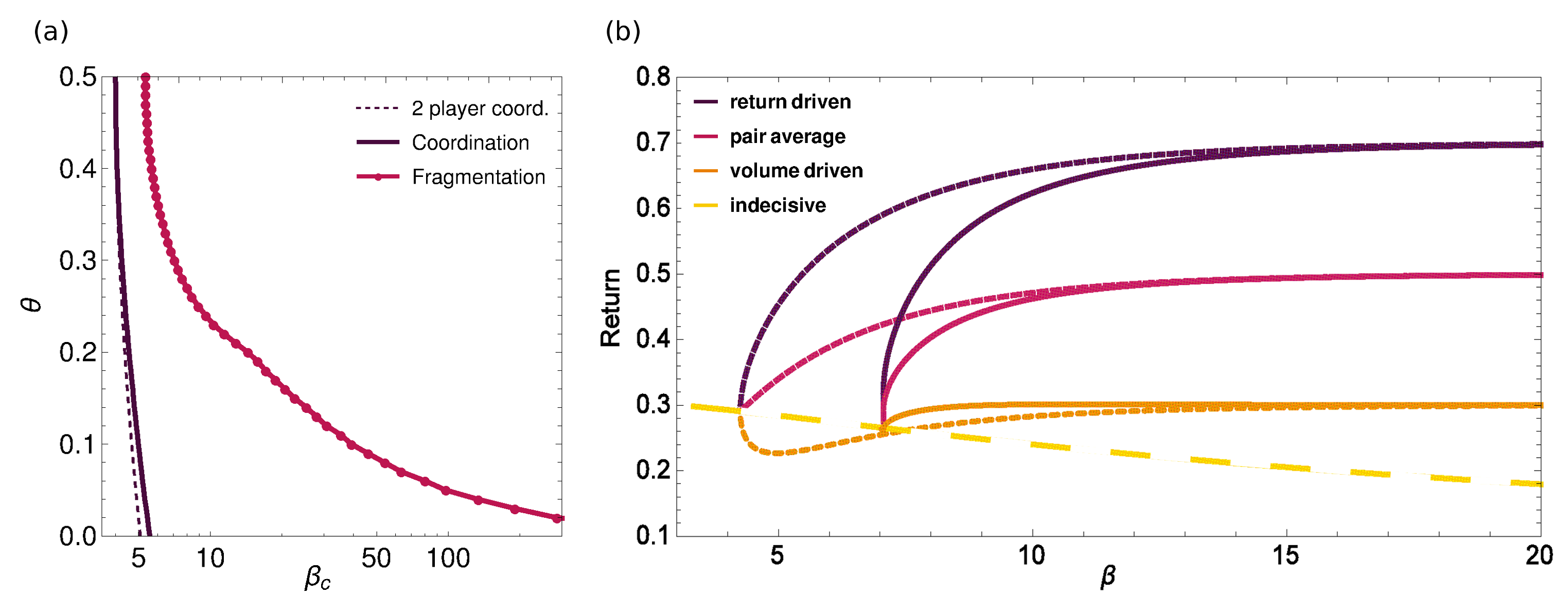}
\caption[Four traders: $(\theta,\beta)$-phase diagram and returns]{Four agents (two buyers and two sellers): phase diagram and returns. (a) Phase diagram showing steady states as a function of intensity of choice $\beta$ and market bias $\theta$. Coordinated steady states exist to the right of the dark violet, solid line and fragmented steady states to the right of the pink solid line with markers. 
(b) Returns against intensity of choice $\beta$ for all agents and separately for return and volume driven agents; market biases are $(\theta_1,\theta_\mo)=(0.3,0.7)$. Dashed lines show coordinated states and solid lines fragmented states.
The yellow line shows the average return in the uncoordinated steady state, continued (dashed) into the instability region at high $\beta$.}
\label{4playersPD}
\end{figure*}

The two player system studied above already exhibited an interesting collective phenomenon -- coordination at a market to enable more trades, sometimes even to the detriment of an individual agent. Turning to fragmentation, where otherwise homogeneous agents nonetheless learn to adopt different policies, the minimal system size where we can expect a similar effect is $N=4$. We first study two identical buyers and two identical sellers, choosing agents with deterministic buy/sell behaviour ($p_{\mathcal B}^i=0$ or 1) for simplicity. A system with four agents is small enough so that we can still easily write down deterministic equations for the evolution of market attractions, but large enough for the first signals of fragmented states to appear as agents can split across the markets in pairs. We consider again
symmetrically biased markets, $\theta_1=1-\theta_{\mo}=\theta$. As before, the market choice behaviour of each agent is determined by his/her market attraction difference $\Delta^{g,i}$.  Here the index $g$ denotes the agent group (buyers or sellers) while $i$ labels agents within each group. For small $r$ the attraction differences again obey deterministic time evolution equations that can be derived by following the reasoning in the previous subsection. The only difference lies in the calculation of the return $S_m^{g,i}(n)$ at a chosen market, which now depends on the choices made by all other players:
\begin{align*}
S^{g,i}_m(n)&=\delta_{m^{g,i}(n),m}\\
&\Big\lbrace\frac{\Sigma_m^g}{2}\delta_{m^{g,-i}(n),m}\left(\delta_{m^{-g,1}(n),m}+\delta_{m^{-g,\mo}(n),m}\right)\\
&+\Sigma_m^g\left(1-\delta_{m^{g,-i}(n),m}\right)\Big(\delta_{m^{-g,1}(n),m}\\
&+\delta_{m^{-g,\mo}(n),m}-\delta_{m^{-g,1}(n),m}\delta_{m^{-g,\mo}(n),m}\Big)\Big\rbrace\ ,
\end{align*}
In the expression above, $\Sigma_m^g$ denotes the deterministic part of the return, which only depends on the chosen market $m$ and the agent type $g$, by analogy with the two player case in Eq.~(\ref{deterministiscore}). The Kronecker $\delta$-symbols ensure other agents are present at the same market $m$. The first term describes the situation where both agents of the same type go to a single market $m$; the return is then zero if no agents of the opposite type are at the same market, $\Sigma^g_m$ if there are two, and $\Sigma^g_m/2$ if there is only one (as our chosen agent then only has probability $1/2$ of being allowed to trade).
On the other hand, when the second player of the same type is not at the same market, the player receives the full return if there is at least one trader of the opposite group present. This is described by the second term. The deterministic equations for $r\to 0$ then take the form
\begin{align*}
\partial_t\Delta^{g,i}(t)=-\Delta^{g,i}(t)+\sum_{m=\mo}^1 m
S_m^{g,i}(t)\ ,
\end{align*}
where $S_m^{g,i}(t)$ has the meaning of returns averaged over a long time window so that the Kronecker deltas in $S^{g,i}_m(n)$ are replaced with their expected values, exactly as in the derivation for two players. We solve these equations numerically and find that for low and intermediate intensity of choice $\beta$ the behaviour is analogous to that for $N=2$, showing a transition from a single uncoordinated fixed point to two coordinated fixed points as $\beta$ increases; throughout this range the agents within each group have identical market attractions.
The novel feature of the $N=4$ system is that, when $\beta$ is increased yet further, four new stable states appear. We call these fragmented as each group of agents now ``fragments'' into two individuals with distinct -- and essentially opposite -- market preferences. Both markets are then populated by a pair of traders, one from each group. The fragmented fixed points appear in stable/unstable pairs and for high enough value of $\beta$ unstable fragmented fixed points become partially fragmented, e.g.\ only one group splits across the markets, while the other group specialize for one market. As these fixed points are not stable we do not show this transition line in Figure~\ref{4playersPD}.

In Figure~\ref{4playersPD} we show the two critical $\beta$ lines (the coordination and the fragmentation threshold) as a function of the market bias $\theta$, for the above scenario of four players with strict buy/sell roles. The coordination line is very close to the one for two players, which is included for comparison. Both the coordination and fragmentation lines follow the same trend, with the threshold in $\beta$ increasing as $\theta$ departs from $0.5$.

The right panel of Figure~\ref{4playersPD} shows return lines for different intensities of choice $\beta$: dashed lines correspond to coordinated states, while solid lines are averages in the fragmented state. Note that the difference in returns in the coordinate state is between {\em groups} of agents, with all agents in a group either return driven or volume driven. In the fragmented states there is one return driven and one volume driven agent in each group, on the other hand.
We note that in the large $\beta$ limit the returns achieved in coordinated and fragmented states become identical. This is because with either pattern of market choices, if these choices are made deterministically then all agents are guaranteed to be able to trade. 
For finite $\beta$, returns in the coordinated state are generally higher than for fragmentation.

Note that the four fragmented states arise because in each agent group there are two ways for assigning the two agents to the two markets. For $N$ agents, one therefore expects $\{(N/2)!/[(N/4)!]^2\}^2$ such states.
This number grows very rapidly with $N$ while the number of coordinated states remains at two.

\begin{figure}[h!]
\centering
\includegraphics[width=0.48\textwidth]{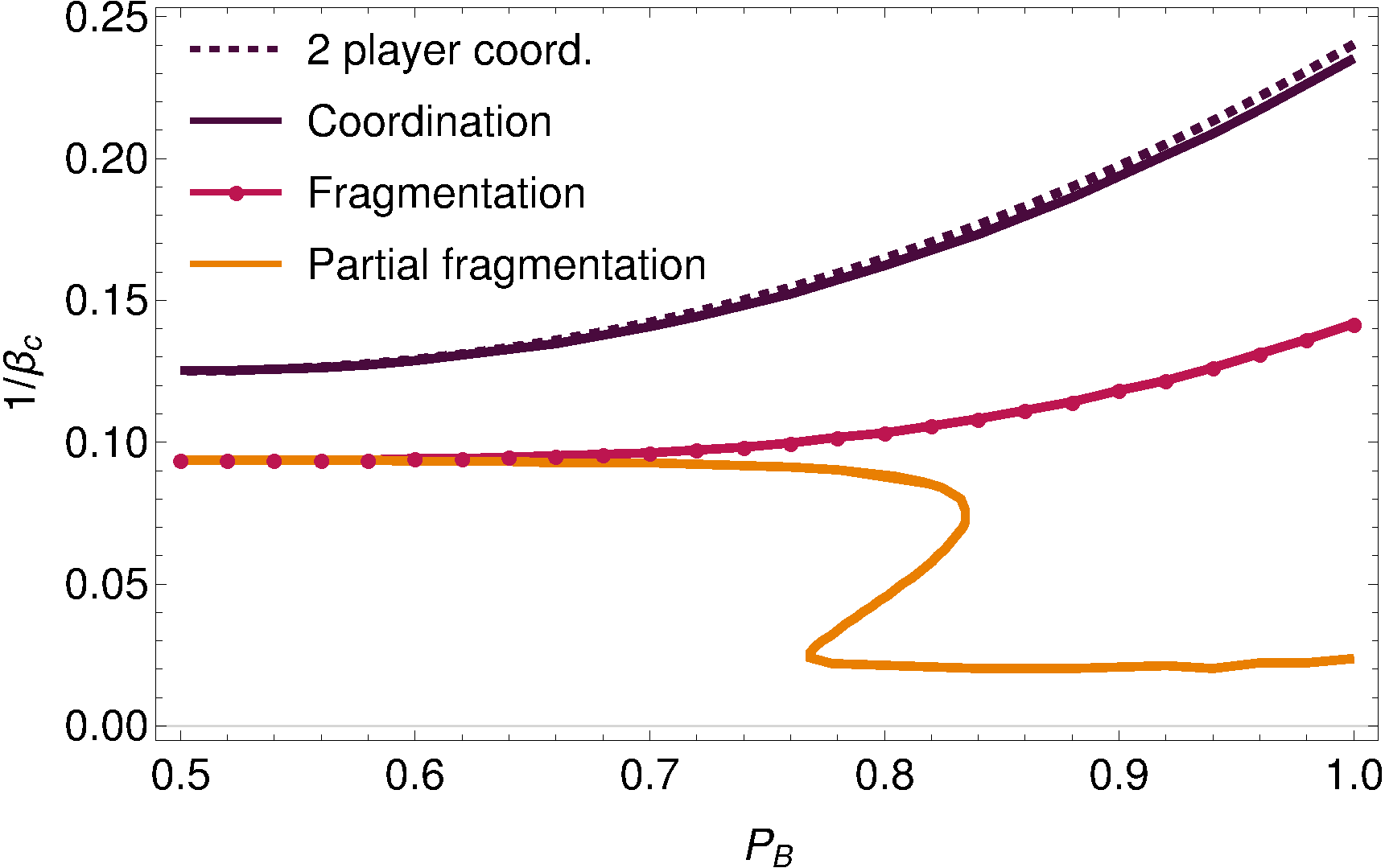}
\caption[Four players $(\beta,p_\mathcal{B})$ phase diagram]{Four traders: phase diagram when buy and sell roles are probabilistic. Coordination takes place below the dark violet (solid) line and fragmentation below the pink line (solid line with circles). These regions shrink when the difference in the buy and sell preferences of the agents decreases ($p_{\mathcal{B}}\to 0.5$), similarly to the trend in the two player system (dashed line). Below the orange line ``partially fragmented'' fixed points exist, where one of the four agents has a preference for the opposite market.}
\label{4playersPBdiag}
\end{figure}

Finally, as in the analysis of the two agent system we can generalise the results by allowing agents to assume the role of buyer with some group dependent probability $p_\mathcal{B}^{(g)}$. We again take these probabilities as symmetric between groups, $p_\mathcal{B}^{1}=1-p_\mathcal{B}^{\mo}=p_\mathcal{B}$.
The deterministic part $\Sigma^g_m$ of the agents' returns is then modified in a manner directly analogous to Eq.~(\ref{deterministiscore}), and one can determine the effect on the existence of the various steady states. 
Figure~\ref{4playersPBdiag} shows the results for the symmetric markets $(\theta_1,\theta_\mo)=(0.3,0.7)$ and symmetric  groups.
As in the system with only two agents, when the traders' preferences for buying are similar ($p_{\mathcal B}\approx 1/2$) they have a weaker incentive to coordinate, resulting in a higher coordination threshold for $\beta$ (for the sake of clearer visualisation we use $1/\beta$ on the y-axis). The same behaviour is seen also for the fragmentation threshold.

We indicate in Figure~\ref{4playersPBdiag} also the regime where a further type of fixed point exists: partially fragmented states. In these states there is a single agent whose market preference is the opposite of that of the other three players, so that only one agent group is fragmented. These states evolve for high enough $\beta$ out of unstable fragmented states, which themselves appear in pairs with the stable fragmented states at the onset of fragmentation.
As we will see below, partially fragmented states exist in the large population limit too, though in a limited region of parameter space. In the small system here they are unstable. Intuitively this is likely to be due to the smaller number of trades: in the large $\beta$ limit of a partially fragmented fixed point, there will be at most one trade per trading period (only two out of the three agents going to one market will be able to trade) while both fragmented and coordinated states lead to two trades. 

\section{\label{sec:largepopulation}Large population limit}
\subsection{\label{sec:singlepopulation}Population with a fixed buying preference}
In studying systems with a small number of agents we have already encountered a rich phenomenology: coordination of agents at a single market, pairwise fragmentation across two markets and even some mixed states where one group fragments while the other specialises in trading at a single market. We now complement and generalise these results by investigating the possible types of steady state in the large population limit. We start with a simple setting, a population in which all agents have identical preferences for buying $p^i_\mathcal{B}=p_\mathcal{B},\forall i$. The assumption of population homogeneity is a strong one, but these traders still undergo fragmentation for a broad range of parameters, while the system is simpler to analyse. Thus it is a useful prelude to the analysis of a population consisting of two or more groups with distinct buying preferences.

To describe the steady states of such an initially homogeneous agent population we follow the distribution of attraction differences ($\Delta^i=A^i_1-A^i_2$) across the population~\cite{aloric2016}. 
The state of each market $m$ enters via the probability with which buy ($\mathcal{B}$) and sell ($\mathcal{S}$) orders are executed successfully~\cite{aloric2016,aloric2017thesis}:
\begin{equation}
  \begin{split}
 T_{\mathcal{B}m}=\min\left(1,\frac{Q_{\mathcal{S}m}}{Q_{\mathcal{B}m}D_m}\right)
  \end{split}
  \quad
  \begin{split}    T_{\mathcal{S}m}=\min\left(1,\frac{Q_{\mathcal{B}m}D_m}{Q_{\mathcal{S}m}}\right).
  \end{split}
\end{equation}
Here the factors $Q_{\mathcal{B,S}m}$ are the probabilities for submitted buy/sell orders to be valid, i.e.\ on the right side of the trading price; the explicit expressions are given in Appendix~\ref{FPappendix}). Note that whereas for small systems we simplified to deterministic order prices, we return here to the full model where bids, $b$, and asks, $a$, are stochastic and the trading price is calculated as in Eq.~\ref{price}. (As explained in Sec.~\ref{sec:model} we choose Gaussian distributions for bids and asks, $a \sim \mathcal{N}(\mu_a,\sigma_a^2)$ and $b \sim \mathcal{N}(\mu_b,\sigma_b^2)$; for numerical evaluations we set $\mu_b-\mu_a=1$, $\sigma_a=\sigma_b=1$) The $D_m$ are demand-to-supply ratios, defined as the number of buyers over number of sellers at market $m$.
For small $r$, the attraction difference distribution evolves according to a Fokker-Planck equation
\begin{align}
\partial_t P(\Delta|p_\mathcal{B},T_\gamma)=
&-\partial_\Delta\left[M_1(\Delta|p_\mathcal{B},T_\gamma)P(\Delta|p_\mathcal{B},T_\gamma)\right]
\label{FokkerPlanck}
\\
&+\frac{r}{2}\partial_\Delta^2\left[M_2(\Delta|p_\mathcal{B},T_\gamma)P(\Delta|p_\mathcal{B},T_\gamma)\right]\ ,
\nonumber
\end{align}
where the drift $M_1$ and diffusion $M_2$ both depend on the buying preference of the agents and on the four trading probabilities $T_\gamma$. (We use $\gamma=(\tau,m)$ as the generic label for a combination of order type $\tau=\mathcal{B},\mathcal{S}$ and market choice $m$.)
The drift term is (see Appendix~\ref{FPappendix} for details and for the explicit expression of the return distribution from which $\langle S_\gamma\rangle$ is calculated)
\begin{align}
&M_1(\Delta|p_\mathcal{B},T_\gamma)\nonumber\\
&=\sum_{m=\mo}^1\sum_{\tau\in\{\mathcal{B},\mathcal{S}\}} m p_\tau T_{\tau m}\langle S_{\tau m}\rangle\sigma_\beta\left(m\Delta\right)
-\Delta\ ,
\label{firstmoment}
\end{align}
where the sum runs over markets $m$ and order types $\tau$ and we use the convention
$p_\mathcal{S}=1-p_\mathcal{B}$. 
The strength of the diffusion term is 
\begin{align}
\label{secondmoment}
&M_2(\Delta|p_\mathcal{B},T_\gamma)=\Delta^2 + \\
&\sum_{m=\mo}^1 \sum_{\tau\in\{\mathcal{B},\mathcal{S}\}}
\Big[
p_\tau T_{\tau m}
(\langle S^2_{\tau m}\rangle-2m\Delta\langle S_{\tau m}\rangle)
\Big]
\sigma_\beta\left(m\Delta\right)\ .\nonumber
\end{align}
The steady state of the Fokker-Planck equation is (see e.g.~\cite{vanKampen}):
\begin{align}
P(\Delta|p_\mathcal{B},T_\gamma)\propto \frac{1}{M_2(\Delta|p_\mathcal{B},T_\gamma)}
\exp\left(-\frac{f(\Delta)}{r}\right)\ ,
\label{stationarydistribution}
\end{align}
where
\begin{align}
f(\Delta)=-2\int_0^\Delta d\Delta'\frac{M_1(\Delta'|p_\mathcal{B},D_m)}{M_2(\Delta'|p_\mathcal{B},D_m)}
\label{freeenergy}
\end{align}
plays a role analogous to a free energy in thermodynamics. When $f(\Delta)$ has a single minimum, $P(\Delta)$ will approach a narrow peak at this location for $r\to 0$ and we have an unfragmented state. Otherwise as many peaks as there are local minima in $f(\Delta)$ will appear, corresponding to a fragmented state: each peak represents a subgroup of agents following a distinct market choice strategy.

Note that in the Fokker-Planck description, the market order parameters $D_m$ that determine the trading probabilities $T_\gamma$ have to be calculated {\em self-consistently} from $P(\Delta)$ \cite{aloric2016,aloric2017thesis}. The same self-consistency condition then also needs to hold at a steady state. Initially we will treat the order parameters as fixed exogenously, however.
Such a situation could arise if, for example, our agents are just a very small fraction of the overall trading cohort, with the latter fixing the demand-to-supply ratio.

\subsubsection*{Fragmentation for $r\rightarrow 0$}

\begin{figure}[h!]
\centering
\includegraphics[width=0.5\textwidth]{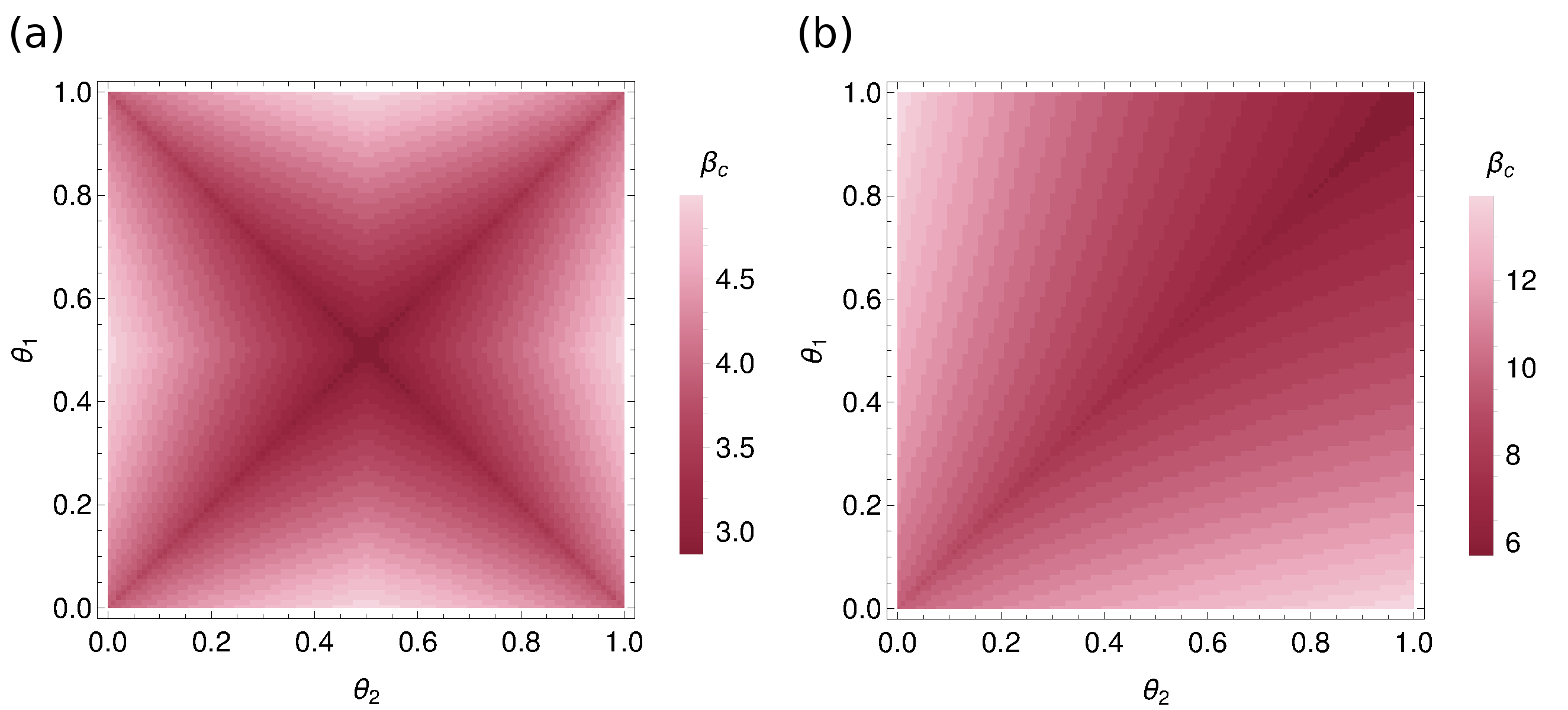}
\caption[Homogeneous population $\beta_c$ as function of $(\theta_1,\theta_\mo)$]{Critical intensities of choice as a function of market biases for (a) an indecisive population ($p_\mathcal{B}=0.5$) and (b) a decisive population of buyers ($p_\mathcal{B}=0.8$).}
\label{singlePopCriticalT}
\end{figure}

In Figure~\ref{singlePopCriticalT} we show how the treshold value of the intensity of choice depends on the market biases $(\theta_1,\theta_\mo)$ for different agent populations, one ``indecisive'' ($p_\mathcal{B}=0.5$) and one made up of ``decisive'' buyers ($p_\mathcal{B}=0.8$); for this calculation we set the order parameters to their ``endogenous'' value following the self consistent procedure outlined in~(\ref{orderparamDef}). We see that for every pair of market biases there is a finite threshold $\beta_c$ above which fragmentation sets in.
When agents are indecisive with regards to buying and selling, the region where fragmentation occurs is greatest when markets are identical or symmetrically biased. For decisive buyers ($p_\mathcal{B}=0.8$), on the other hand, the fragmentation threshold is lowest when the markets are identical. Intermediate values of $p_\mathcal{B}$ provide a smooth interpolation between these two situations.

\begin{figure}[h!]
\includegraphics[width=0.5\textwidth]{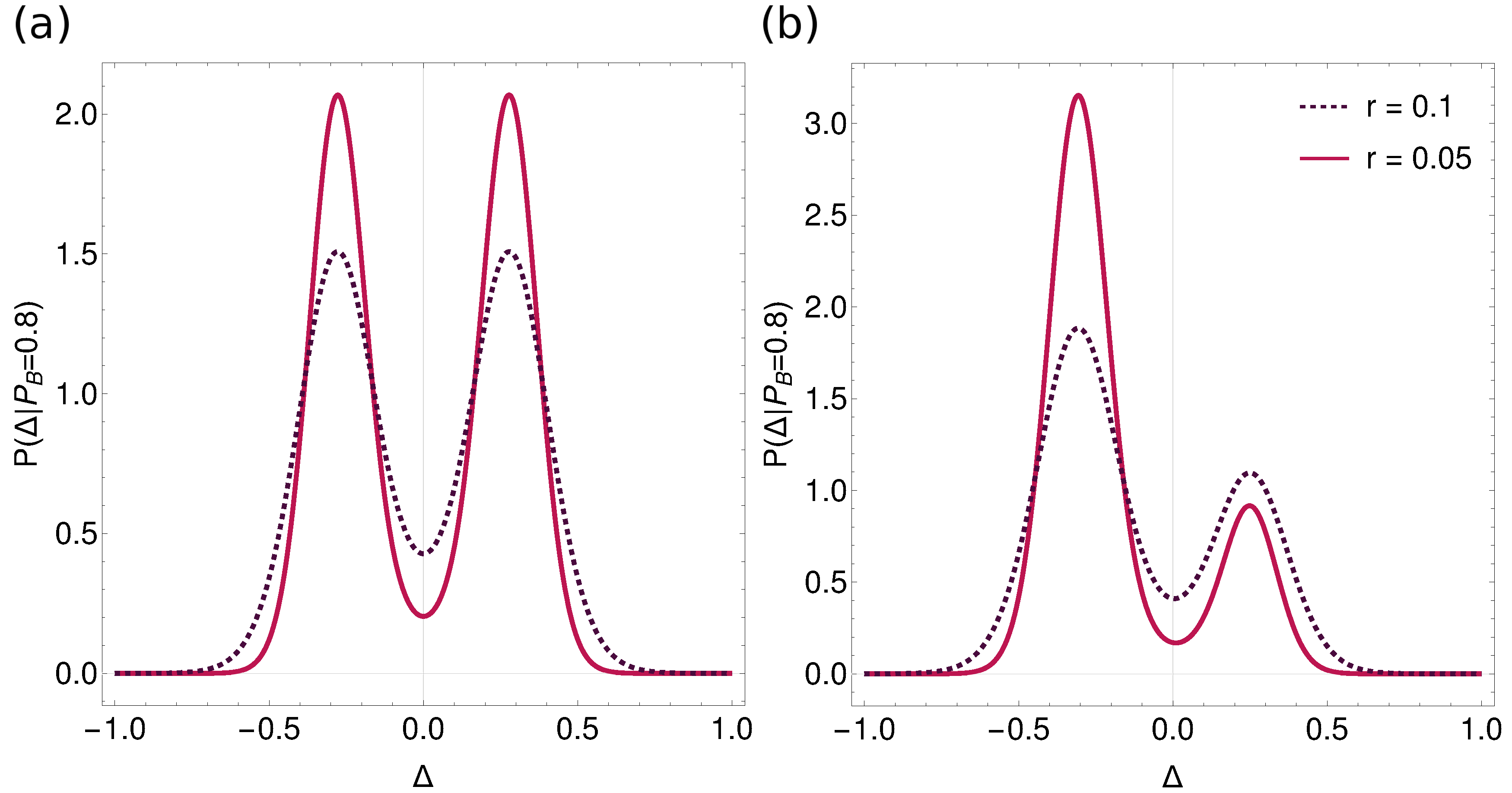}
\caption[Steady states of decisive buyers]{Steady-state attraction difference distributions of decisive buyers ($p_{\mathcal{B}}=0.8$). We compare steady states at $\beta=20$ for (a) two unbiased markets $(\theta_1,\theta_\mo)=(0.5,0.5)$ and (b) two symmetrically biased markets $(\theta_1,\theta_\mo)=(0.3,0.7)$, for $r=0.1$ (dashed dark violet line) and 0.05 (solid pink line). The distributions are strongly and weakly fragmented, respectively: on the right, the relative height of the lower peak decreases as $r$ is reduced.}
\label{differentRsol}
\end{figure}

To understand more closely the properties of the fragmented states, we show 
in Fig.~\ref{differentRsol} the steady state distributions of traders with $p_\mathcal{B}=0.8$ when faced with the choice between two unbiased or two symmetrically biased markets. To understand the trend with $r$, we show the distributions for two different values of $r$ in each case.
As expected from (\ref{stationarydistribution}) the peak width decreases as $\approx\sqrt{r}$ with decreasing $r$, but in Figure~\ref{differentRsol} (b) we see that the relative peak heights also depend on $r$.
In fact, if the peaks are located at attraction differences $\Delta_1$ and $\Delta_2$ then the peak height ratio can be written as 
\begin{align}
&\frac{P(\Delta_1|p_\mathcal{B},T_\gamma)}{P(\Delta_2|p_\mathcal{B},T_\gamma)}\nonumber\\
&=
\frac{M_2(\Delta_2|p_\mathcal{B},T_\gamma)}{M_2(\Delta_1|p_\mathcal{B},T_\gamma)}\exp\left(-\frac{f(\Delta_2)-f(\Delta_1)}{r}\right)\ .
\label{weightratio}
\end{align}
This ratio can stay finite for $r\to 0$ only when
\begin{align}
f(\Delta_1)=f(\Delta_2),
\end{align}
and we call this situation \textit{strong} (S) fragmentation as it survives even in the $r\rightarrow 0$ limit.
This is the situation in Figure~\ref{differentRsol} (a).
If the free energies at the two peaks are unequal, on the other hand, one continues to have two peaks in $P(\Delta)$ for any nonzero $r$ but the height of one peak decreases (exponentially in $1/r$) as $r$ goes to zero. We call this behaviour, which is illustrated in Fig.~\ref{differentRsol} (b), \textit{weak} (W) fragmentation because the lower peak may become unobservably small for low $r$; in the strict limit $r\to 0$, the distribution $P(\Delta)$ becomes unimodal again. The strong-weak distinction as defined applies  literally only to this $r\to 0$ limit; at nonzero $r$ it becomes a crossover between fragmented states where the emergent subgroups have roughly even (strong) or very different (weak) sizes. At the weakly fragmented state most of the trades happen at a single market (increasingly so as $r$ decreases) we relate this state to market consolidation, thus the question of fragmentation versus consolidation becomes question of strong versus weak fragmentation in our set up.
\begin{figure*}[ht!]
\includegraphics[width=0.9\textwidth]{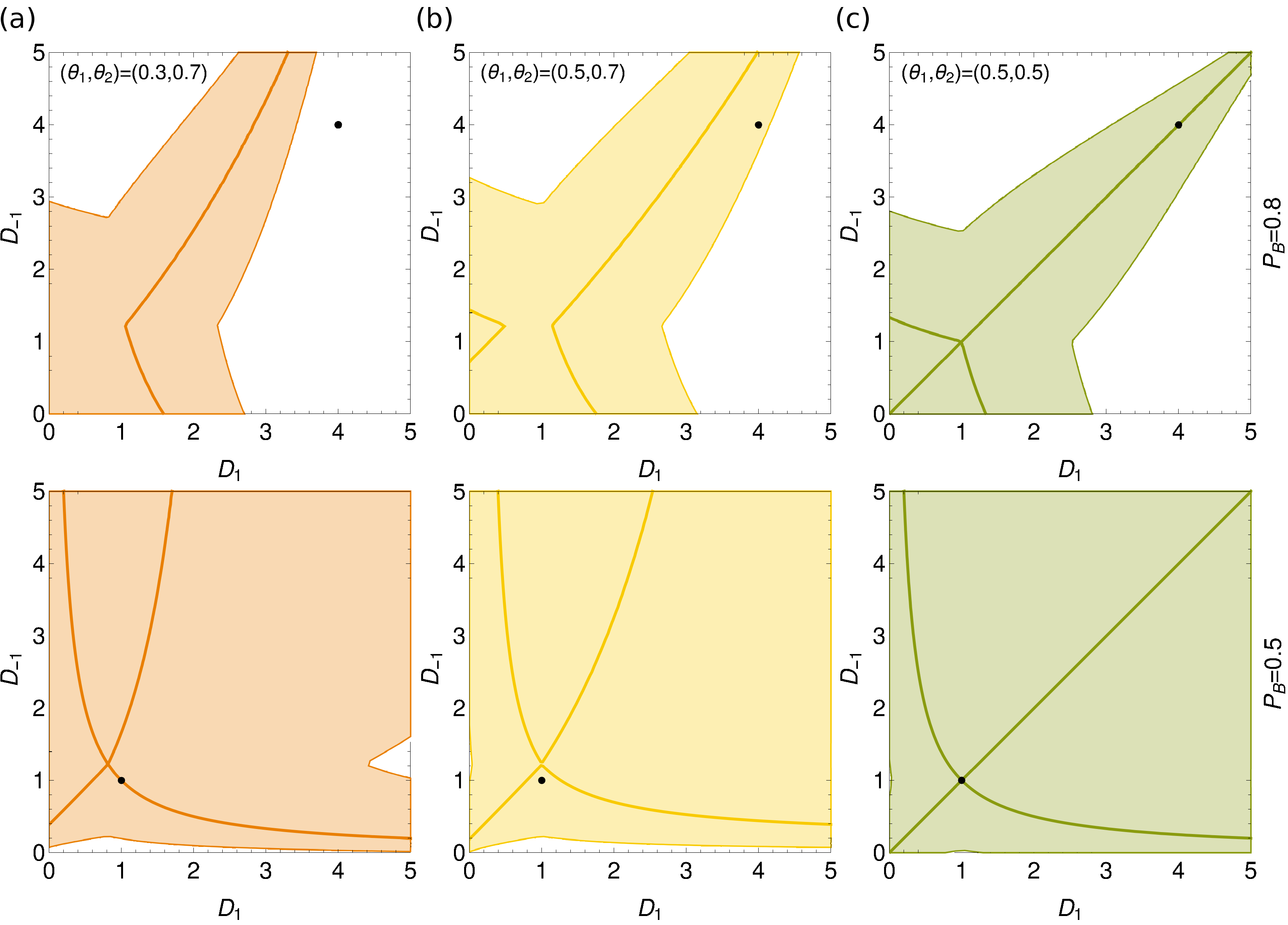}
\caption[Single population: steady state types in the space of market order parameters $(D_1,D_\mo)$]{Single population: steady-state types in the space of market order parameters $(D_1,D_\mo)$ for $\beta=8.5$. Shown on top is the population of decisive buyers ($p_\mathcal{B}=0.8$) and on bottom the
indecisive population ($p_\mathcal{B}=0.5$) for (a) symmetrically biased markets $(\theta_1,\theta_\mo)=(0.3,0.7)$, (b) one unbiased and one biased market $(\theta_1,\theta_\mo)=(0.5,0.7)$, and (c) two unbiased markets $(\theta_1,\theta_\mo)=(0.5,0.5)$. Coloured regions indicate where weakly fragmented states exist (for $r\rightarrow 0$). Solid lines inside these regions indicate strongly fragmented states.
}
\label{segregationCondition}
\end{figure*}

Now that we have a method for finding steady states and classifying them we return to the space of market order parameters and investigate where fragmentation occurs. In Figure~\ref{segregationCondition} we show where weakly (coloured region) and strongly (solid lines inside these regions) fragmented states appear,
at a fixed intensity of choice $\beta=8.5$. We compare again indecisive ($p_B=0.5$) and decisive buyers ($p_B=0.8$), for three different market setups.
We first note that the weak fragmentation region encompasses a very wide range of market conditions (order parameters $D_m$) for indecisive buyers, but shrinks significantly when the agents have stronger preferences for buying. 
Looking at the dependence on market setup, an obvious feature is that for two unbiased markets (shown in (c)), equal  demand-to-supply ratios ($D_1=D_\mo$ line) produce strong fragmentation for both types of agents. This makes sense as the markets are then identical both in their setup $\theta_1=\theta_\mo$ and in the prevailing market conditions, making it easy for groups of agents with opposite market preferences to coexist.
For the indecisive agents who will act as buyers or sellers with equal probability, the same situation arises when the markets have exactly opposite demand-to-supply ratios ($D_1=1/D_\mo$) and therefore still offer them identical average returns.

With increasing market biases (Figure~\ref{segregationCondition}, (a) and (b)) the picture obtained for two unbiased markets changes largely smoothly, though note that for decisive buyers (top row) the two crossing lines of strong fragmentation detach into two separate lines ((b) top), with one eventually disappearing out of range.

\paragraph*{Market order parameters.} 
So far we have looked at fragmentation behaviour driven by exogenously set market conditions (demand-to-supply ratios). We now return to our model as originally set out, where only the adaptive agents we describe trade at the two markets. This leads to the question: will a population endogenously create market conditions needed for its fragmentation? 

For the case of traders with homogeneous buying preferences $p_\mathcal{B}$ this question can be answered relatively straightforwardly. If the steady state distribution of attraction differences is $P(\Delta|p_\mathcal{B},T_\gamma)$, then the fractions of the whole population buying and selling at market $m$ are:
\begin{align}
N_{\mathcal{B} m}&=p_\mathcal{B}\int d\Delta P(\Delta|p_\mathcal{B},T_\gamma) \sigma_\beta(m\Delta)\ ,\nonumber\\
N_{\mathcal{S} m}&=(1-p_\mathcal{B})\int d\Delta P(\Delta|p_\mathcal{B},T_\gamma) \sigma_\beta(m\Delta)\ .
\label{buysellnumber}
\end{align}
The demand-to-supply ratio then does not in fact depend on the market preference distribution
\begin{align}
D_m=\frac{N_{\mathcal{B}m}}{N_{\mathcal{S}m}}
=\frac{p_\mathcal{B}}{1-p_\mathcal{B}}
\label{orderparamDef}
\end{align}
and is fully determined by $p_B$. In the space of market order parameters in Figure~\ref{segregationCondition}, this endogenous set of market conditions is marked with a black dot. We see that -- at high enough $\beta$ -- the population of indecisive buyers fragments strongly when the markets are unbiased or symmetrically biased, and one can check that these results hold independently of the specific market biases used in the figure.
Decisive buyers, on the other hand, fragment strongly only if the markets are equal (the figure shows only $\theta=0.5$ but the statement is true for general $\theta$). Otherwise weak fragmentation occurs, although -- see Figure~\ref{segregationCondition} (top (c)) -- when the markets are very different at a $\beta$ above that used in Figure~\ref{segregationCondition}.

With these insights it is worth revisiting Fig.~\ref{singlePopCriticalT}. It shows the existence of the fragmentation threshold $\beta_c$ for all market biases, and we recall that this threshold is defined as the point where $P(\Delta)$ first acquires two peaks. From what we have seen above we now understand that
for most combinations of market biases, the steady state one finds above $\beta_c$ is a weakly fragmented one. The exceptions are the dark lines in Fig.~\ref{singlePopCriticalT}, which indicate equal ($\theta_1=\theta_\mo$) or symmetrically biased ($\theta_1=1-\theta_\mo$) markets.

\section{\label{sec:2populations}Two group population}

So far in our analysis of the large size limit of a homogeneous population of traders with buying preference $p_\mathcal{B}$, we showed how for any given pair of market order parameters $(D_1, D_\mo)$ we can determine the population steady state. We identified three possible types of steady states: unfragmented (U), weakly fragmented (W) and strongly fragmented (S). 
We now generalise the investigation to populations of agents consisting of groups with different buying preferences.
We demonstrate the approach for the case of two groups of the same size, but the principles are general and can be extended to larger numbers of groups or different group sizes. We denote a steady state of a population consisting of two groups with a pair of letters (X,X'). Here X,X' $\in\{$U,W,S$\}$ indicates the type of steady state for each group, producing nine different types of population steady states.

We can now find, in the space of market order parameters $(D_1,D_\mo)$, the  domains of different state types as we did in Figure~\ref{segregationCondition}. We can use the figure directly to read off the steady states at $\beta=8.5$ of a population of two groups with buying preferences $(p_\mathcal{B}^{(1)}, p_\mathcal{B}^{(2)})=(0.8,0.5)$. For example, when the market order parameters are $(D_1,D_\mo)=(5,5)$ (the top right corner of all the diagrams), the steady state of the two group-population is (U,W) when the markets are symmetrically biased or biased/unbiased (Fig.~\ref{segregationCondition} left and centre) and (S,S) when both markets are unbiased (right diagrams). This simple analysis can be extended to any number of groups because, for market order parameters that are fixed exogeneously, the groups are independent.

Our primary interest, however, lies in the case of endogenous market conditions where the agents we model capture the entire trading population and thus define their own market order parameters. In this case, we need to find the steady states self-consistently. We have previously described a procedure for doing this, for nonzero $r$~\cite{aloric2016}: starting from some initial market order parameters $(D_1,D_\mo)$ one calculates the steady states and updates $(D_1,D_\mo)$ iteratively, converging eventually to a self consistent set of order parameters. Here we aim to get a complete picture of all possible steady states, independently of initial conditions. To do this, we start from the update equation for the market order parameters from the iterative approach. These are simply the definitions of the market order parameters (Eqs.~\ref{buysellnumber},\ref{orderparamDef}) extended to two groups: 
\begin{widetext}
\begin{align}
\label{Eq:2populationDm}
D'_m&=\frac{N_{\mathcal{B}m}^{(1)}+N_{\mathcal{B}m}^{(2)}}{N_{\mathcal{S}m}^{(1)}+N_{\mathcal{S}m}^{(2)}}=\frac{p_\mathcal{B}^{(1)}\int d\Delta \sigma_\beta(m\Delta) P(\Delta|p_\mathcal{B}^{(1)},T_\gamma)+p_\mathcal{B}^{(2)}\int d\Delta \sigma_\beta(m\Delta) P(\Delta|p_\mathcal{B}^{(2)},T_\gamma)}{(1-p_\mathcal{B}^{(1)})\int d\Delta \sigma_\beta(m\Delta) P(\Delta|p_\mathcal{B}^{(1)},T_\gamma)+(1-p_\mathcal{B}^{(2)})\int d\Delta \sigma_\beta(m\Delta) P(\Delta|p_\mathcal{B}^{(2)},T_\gamma)}\ .
\end{align}
\end{widetext}
We can now define, in the market order parameter space, the two loci where $D_1'=D_1$ and $D_\mo'=D_\mo$, respectively, meaning that one of the order parameters is already self-consistent. The intersection of these loci (two lines, for our case of two markets) then gives us all the self-consistent sets of market order parameters.
To distinguish weak and strong fragmentation the limit $r\to 0$ ought to be taken. To avoid numerical issues we use here instead a small nonzero $r$ to determine the attraction difference distributions $P(\Delta|p_\mathcal{B}^{(g)},T_\gamma)$ from which the $D_m'$ are calculated.
In most of what follows we focus on symmetric market setups ($\theta_1=1-\theta_\mo$) and symmetric agent buying preferences ($p_\mathcal{B}^{(1)}=1-p_\mathcal{B}^{(2)}=p_\mathcal{B}$). To avoid having too many parameters to vary we will fix the market biases to the default values ($\theta_1,\theta_\mo$)=(0.3,0.7) unless stated otherwise.

\subsection{Transitions in populations of decisive and indecisive traders.}

As shown in previous sections, the intensity of choice $\beta$ is a crucial parameter determining whether the steady state in a system is fragmented or consolidated. Here we build upon this analysis by investigating how the nature of steady states changes as $\beta$ is increased. 
We start this section with examples of steady states of a population with ``decisive'' traders $(p_\mathcal{B}^{(1)},p_\mathcal{B}^{(2)})=(0.8,0.2)$ as well as one with largely ``indecisive'' traders $(p_\mathcal{B}^{(1)},p_\mathcal{B}^{(2)})=(0.55,0.45)$. We then generalize these results to a full phase diagram for the $r\rightarrow 0$ limit, giving the number and type of steady states as a function of the intensity of choice $\beta$ and the buying preference $p_\mathcal{B}$. 

\subsubsection{Decisive traders.} 

\begin{figure*}[!ht]
\includegraphics[width=\textwidth]{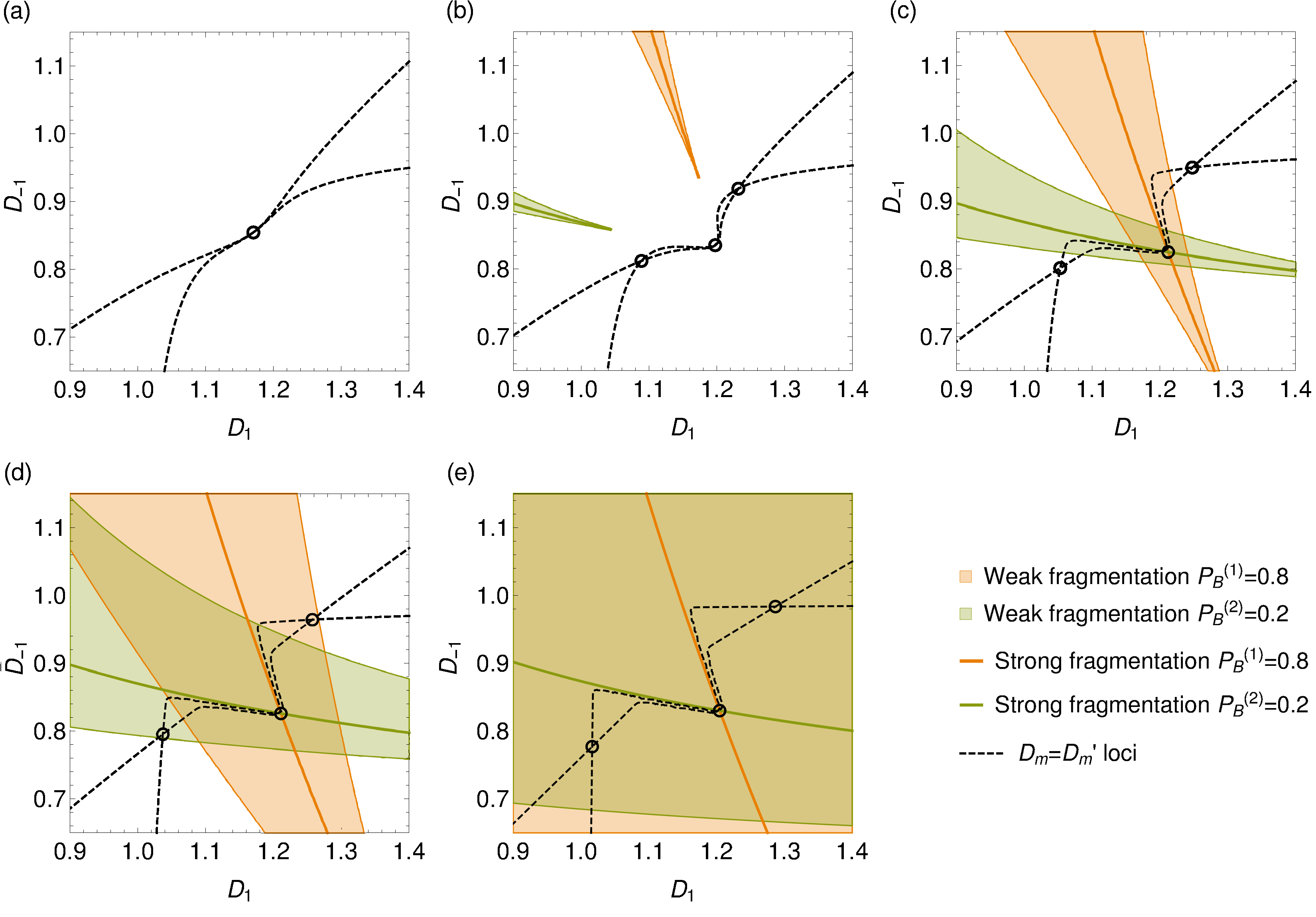}
\caption[Steady states of population with decisive traders]{Steady states of traders with decisive buy and sell preferences: order parameter diagrams. For a two-group system with $(p_\mathcal{B}^{(1)},p_\mathcal{B}^{(2)})=(0.8,0.2)$, each diagram shows the order parameter self-consistency lines (dashed black) and for each group the weak fragmentation region and the strong fragmentation line; $r=0.001$ throughout.
(a) Single (U,U) solution with $\beta=1/0.31$, (b) three (U,U) solutions with $\beta=1/0.29$, (c) one (S,S) and two (U,U) solutions with $\beta=1/0.265$, (d) (U,W), (S,S) and (W,U) solutions with  $\beta=1/0.245$ and (e) (W,W), (S,S), and (W,W) solutions with  $\beta=1/0.2$. We use the following abbreviations for the steady state of each group: U, unfragmented; W, weakly fragmented; and  S, strongly fragmented.}
\label{pb08transitions}
\end{figure*}

In Figure~\ref{pb08transitions} we show, for a series of different $\beta$, the market order parameter space $(D_1,D_\mo)$ with the weak fragmentation region and the strong fragmentation line marked for both groups of a population with $(p_\mathcal{B}^{(1)},p_\mathcal{B}^{(2)})=(0.8,0.2)$. The order parameter self-consistency lines are also shown.

In panel (a) of Figure~\ref{pb08transitions}, we show the low $\beta$ regime ($\beta=1/0.31$), just before the onset of fragmentation. Note that for this $\beta$, the steady states of both groups are unfragmented across the entire range of market order parameters shown. The unique intersection of the $D'_m=D_m$ loci identifies a single steady state of type (U,U).
Panel (b) shows a just slightly increased $\beta=1/0.29$ where most market order parameters settings still give unfragmented states but there are now three intersections of the self-consistency loci, giving as many (U,U) steady states. In the steady state that is a continuation of the low $\beta$ solution the agents show only mild preferences among the markets, with buyers slightly preferring the market that gives higher returns for buyers and similarly for sellers. The other two unfragmented solutions correspond to coordination at one of the two markets so that overall the situation is similar to the one we saw for $N=2$ and $N=4$.

Increasing $\beta$ further (Fig.~\ref{pb08transitions}(c)) one crosses the threshold ($\beta_c \approx 1/0.28$ here) where one of the unfragmented solutions first fragments -- the continuation of the low $\beta$ state is now in the strongly fragmented domain of both groups, while the other two steady states remain unfragmented. Note that since the weak fragmentation regions surround the strong fragemtation lines for both agent groups, there must in fact be a narrow range of slightly lower $\beta$-values where the fragmentation is weak: the low-$\beta$ solution must change from (U,U) through (W,W) to (S,S) as $\beta$ increases. 

In Fig.~\ref{pb08transitions}(d,e) one observes that with increasing $\beta$ the fragmentation regions keep growing. This results in the two unfragmented (U,U) solutions changing first into (U,W) and finally (W,W).

We note that inferences about stability from diagrams like Fig.~\ref{pb08transitions} are in general unwarranted; e.g.\ the initial pitchfork bifurcation from one to three (U,U) states in Fig.~\ref{pb08transitions}(a,b) does not necessarily imply that the middle solution is unstable. It {\em would} be unstable under repeated updating from $D_m$ to $D_m'$. However, Eq.~(\ref{Eq:2populationDm}) shows that this is not the real dynamics but would correspond to a scenario where the dynamics of the order parameters is slowed down artificially so that agents always have time to equilibrate their attraction difference distributions $P(\Delta|p_{\mathcal B}^{(g)},T_\gamma)$ to the current order parameter values.

We highlight one further feature of Fig.~\ref{pb08transitions}: for small $r$ as used in the figure, the order parameter self-consistency lines tend to follow segments of the strong segregation lines before emerging on either side into a weak segregation region. This can be understood by noting that the self-consistency line for e.g.\ $D_1$ is the zero contour of the function $D_1'(D_1,D_{-1})-D_1$ in the order parameter plane. This function varies steeply as a strong segregation line is crossed, developing discontinuities that look like cliff edges for $r\to 0$. A contour line that hits such a cliff must follow the line of the cliff before returning to the smooth parts of the landscape, which is the effect we see in Fig.~\ref{pb08transitions}. 

The cliff edges themselves arise because on a segregation line, the free energy function $f(\Delta)$ in Eq.~(\ref{weightratio}) has two minima of equal height. A small change of $O(r)$ in $D_1$ or $D_{-1}$ will cause similar small changes in the height of these minima, but from Eq.~(\ref{weightratio}) this is enough to cause the weight ratio between the two peaks in $P(\Delta)$ to shift by a factor of order unity. Changes larger than this will transfer all weight from one peak to the other, and correspondingly modify $D_1'$ by a finite amount. For $r\to 0$ the required order parameter changes become infinitesimal, leading to the cliff edge structure of $D_1'-D_1$ and analogously $D_{-1}'-D_{-1}$.

\begin{figure*}[ht!]
\includegraphics[width=\textwidth]{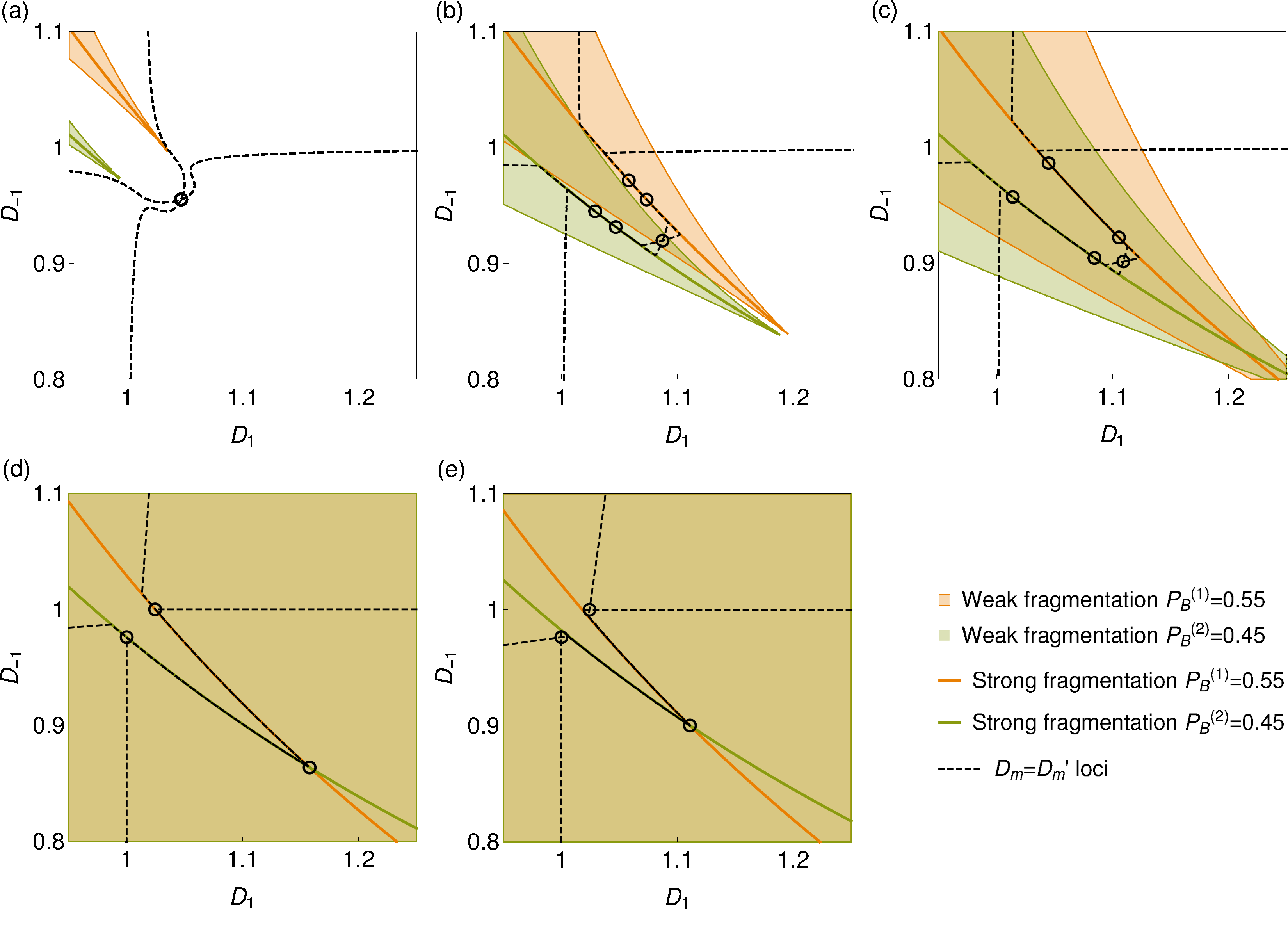}
\caption[Types of steady states of population with indecisive traders]{Steady states of largely indecisive traders: order parameter diagrams. We show the behaviour of largely indecisive traders $(p_\mathcal{B}^{(1)},p_\mathcal{B}^{(2)})=(0.55,0.45)$ for different intensities of choice $\beta$ in the large memory limit, evaluated numerically for $r=0.00001$: (a) $\beta=1/0.31$, one unfragmented solution (U,U); (b) $\beta=1/0.285$, one weakly fragmented (W,W) and four partially fragmented (U,S) states; (c) $\beta=1/0.27$ one weakly (W,W) and four partially fragmented (W,S) states; (d) $\beta=1/0.1$, one strongly fragmented (S,S) and two partially fragmented (W,S) states; and (e) $\beta=1/0.05$, one strongly fragmented (S,S) and two weakly fragmented states (W,W).
}
\label{pb055transitions}
\end{figure*}

\subsubsection{Indecisive traders.}
We now compare the results above with those for a population consisting of two agent groups with only weak preferences for buying and selling, $(p_\mathcal{B}^{(1)},p_\mathcal{B}^{(2)})=(0.55,0.45)$.
The motivation for this comes from the fact that
agents with only mild buy/sell preferences should develop weaker preferences for markets that offer higher returns for buyers or sellers. They will also not be penalized much if only a single group populates a market, as such an arrangement will still sustain a large number of trades.

In Figure~\ref{pb055transitions} we observe several differences compared to the situation in Figure~\ref{pb08transitions} for decisive traders, most notably with regards to the number of solutions. Specifically, as can just be discerned from Figure~\ref{pb055transitions}, on crossing the fragmentation threshold {\em four} new states appear. 
These states are also different in nature: they are partially fragmented in the sense that one group of agents is strongly fragmented and thus retains a bimodal distribution of attraction differences for $r\to 0$ while the other is either weakly fragmented or unfragmented. 
We have seen a similar state in the systems with four agents although there it was unstable because it reduced the number of possible trades.
In the large population limit, having one fragmented and one unfragmented group of agents still leaves many possibilities for trading, especially for indecisive agents where roughly half of each group of agents will probabilistically assume the role of buyer or seller in each trading round. On the general grounds discussed above, the appearance of the partially fragmented (U,S) states is expected to proceed via (U,W) states, though again the $\beta$-range where the latter appear is numerically small.

When the intensity of choice $\beta$ is increased beyond that in Fig~\ref{pb055transitions}(a), the low-$\beta$ solution transitions from (U,U) to (W,W) (panel (b)) and eventually (S,S) (panel (d)), i.e.\ both agent groups fragment first weakly then strongly. Comparing the partially fragmented solutions in panels (b) and (c) we see that they change from (U,S) to (W,S); finally two of them merge with the uncoordinated (W,W) state into an (S,S) state. The other two partially fragmented states eventually transition into (W,W) states; as in the case of decisive traders, these represent coordination of the agents at a single market.

\subsection{$(\beta, p_\mathcal{B})$ Phase Diagram} 
\begin{figure*}[ht!]
\includegraphics[width=0.9\textwidth]{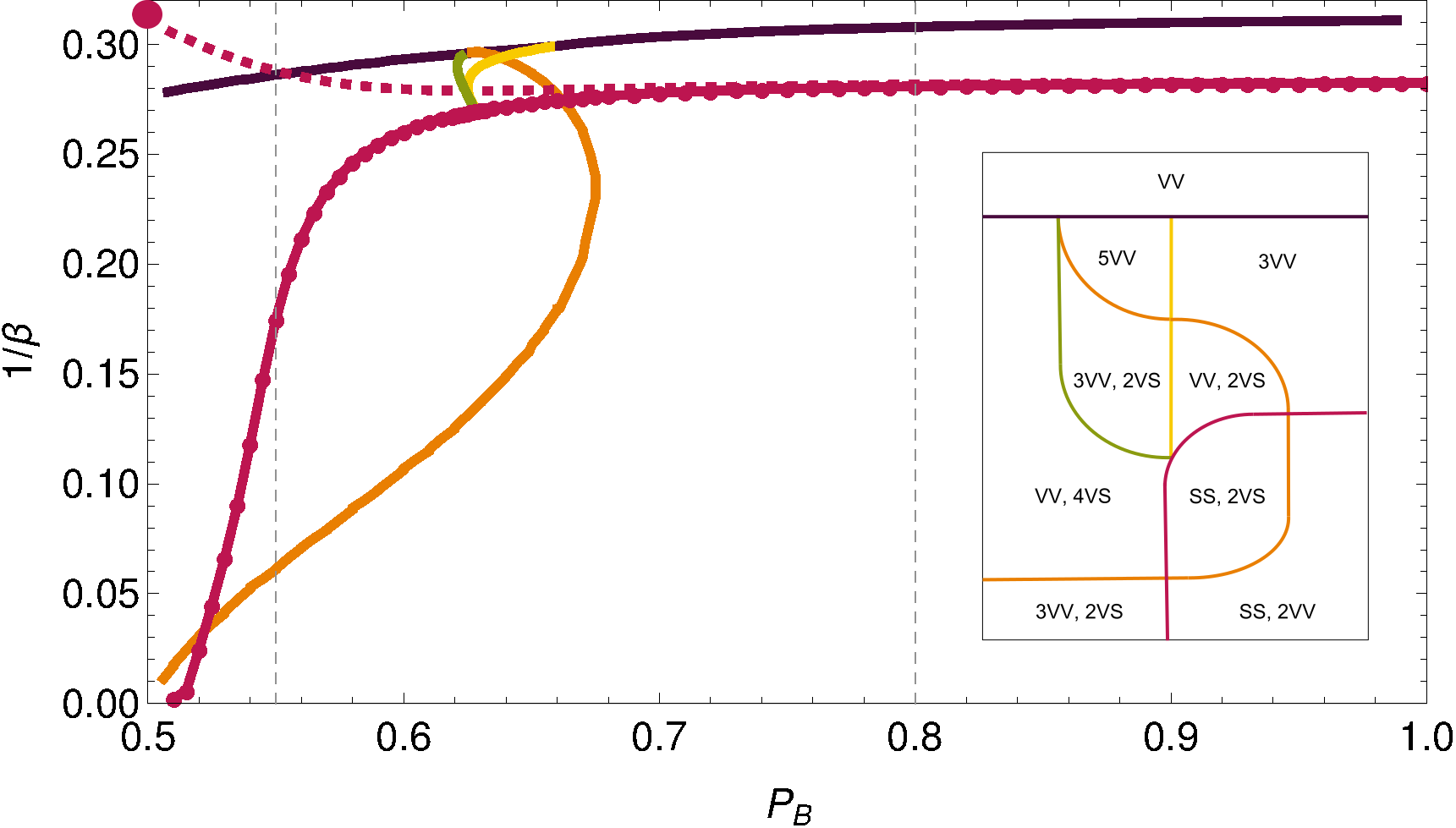}
\caption[Types of steady states for two symmetric agent groups, shown in $(\beta,p_\mathcal{B})$ space]{Types of steady states for two symmetric agent groups, shown in $(\beta,p_\mathcal{B})$ space. Crossing each of the lines changes either the number or the type of steady states. The insert shows the distorted but topologically equivalent diagram to show the phase diagram regions more clearly. The dark violet and yellow lines show the change in solution multiplicity, the pink solid line with circles shows the strong fragmentation line, the pink dashed line shows the weak fragmentation of the uncoordinated low-$\beta$ solution, the orange and green lines show the partial fragmentation. Here V denotes a unimodal distribution of agents' market preferences in the $r\rightarrow 0$ limit, i.e.,\ an unfragmented (U) or weakly fragmented (W) steady state and S denotes a strongly fragmented steady state.}
\label{TpBphaseDiagram}
\end{figure*}

We have observed both market consolidation and fragmentation when a population is faced with a choice of two symmetric markets, depending on the different choice of system parameters ($p_\mathcal{B},\beta$). We next vary these parameters systematically to construct a detailed phase diagram and study the regions where one finds the various states that we described above. The size of these regions then also gives an indication of how typical the different scenarios are. We continue to focus on symmetric markets with $(\theta_1,\theta_\mo)=(0.3,0.7)$ but note that calculations for other (symmetric) market settings give qualitatively similar results. In Figure~\ref{TpBphaseDiagram} we show the phase diagram in the space of intensity of choice $\beta$ and group preference for buying $p_\mathcal{B}\equiv p_\mathcal{B}^{(1)}$. This diagram is the large population analogue of the diagram for four agents ($N=4$) shown in Fig.~\ref{4playersPBdiag}. There we had identified regions with states that are unfragmented and indecisive (low $\beta$), unfragmented and coordinated, fragmented,  and partially fragmented. Broadly these types of states persist in the large population limit, but they have additional structure that makes for a richer phase diagram.

To help visualize the structure of the phase diagram, we show an additional version as an inset that has been distorted to preserve the topology but make even small regions in the phase diagram visible.
Also, to avoid having too many separate regions we do not distinguish in the diagram between unfragmented (U) and weakly fragmented (W) states, which both have distributions of market preferences that become unimodal for $r\to 0$. We label such states collectively V to separate them from strongly fragmented states (S) with their bimodal market preference distributions.
Two vertical dashed lines mark the two scenarios of decisive and indecisive traders studied above (see Figs.~\ref{pb08transitions} and \ref{pb055transitions}).

We now look in more detail at the structure of Figure~\ref{TpBphaseDiagram}.
Crossing any line in the phase diagram either changes the number of population solutions or the nature of the steady state for one or both agent groups. We note that due to the symmetry of the system we consider, many of the changes for the two groups happen simultaneously. In the insert, regions of the parameter space are laid out according to the number of solutions -- five solutions on the left  and three on the right, with a single solution in the small $\beta$ region at the top.

The dark violet line in Figure~\ref{TpBphaseDiagram} is the line where the multiplicity of states changes from one to three or five. Looking at the order parameter self-consistency lines shows that this transition takes place via a pitchfork bifurcation in the former case, and two symmetric saddle-node bifurcations in the latter. The dark violet line is an analogue of the line shown in the same colour in the phase diagram (Fig.~\ref{4playersPBdiag}) of the system with $N=4$ players. The region of multiple solutions has grown for large $N$, but the inverse critical intensity of choice $1/\beta_c$ is still an increasing function of $p_\mathcal{B}$.

As in Fig.~\ref{4playersPBdiag}, the pink line with circles in Figure~\ref{TpBphaseDiagram} marks the appearance of a steady state where both groups are strongly fragmented. We observe that the critical intensity of choice where this happens diverges ($1/\beta\to 0$) as $p_\mathcal{B}\rightarrow 0.5$, i.e.\ the region of strong fragmentation shrinks as the difference between the groups' buying preferences diminishes.
Further lines in the phase diagram show where the solution multiplicity changes directly from three to five (yellow line), and where partial fragmentation occurs (green and orange line) as individual solutions transition from (V,V) to (V,S).
Note that in the large population limit such partially fragmented states appear only for populations with moderate preferences for buying, in contrast to the system with $N=4$ agents (Fig.~\ref{4playersPBdiag}) where they exist for all $p_\mathcal{B}$.

We mark one further line (dashed pink) in the main graph of Figure~\ref{TpBphaseDiagram}, showing the transition within the small-$\beta$ (V,V) solution from the unfragmented (U,U) to the weakly fragmented (W,W) state. With this we make an explicit connection to results reported previously (Figure 7 in \cite{aloric2016}) where we investigated the appearance of (weak) fragmentation with increasing intensity of choice.
We note that for the system of our first case study $p_\mathcal{B}=0.8$ the thresholds for weak and strong fragmentation almost overlap -- the region of the weakly fragmented indecisive state is very narrow for this choice of parameters and in general for $p_\mathcal{B}$ above $\approx 0.7$, while it becomes larger for indecisive traders.
The pink circle on the $y$-axis marks the end of the weak fragmentation line. It turns out that this is the {\em strong} fragmentation threshold of the homogeneous population with even preference for buying and selling ($p_\mathcal{B}=0.5$), with the change from weak to strong fragmentation caused by the additional symmetry between the two groups for this value of $p_\mathcal{B}$. 

Interestingly, there are two distinct regions in the phase diagram of Figure~\ref{TpBphaseDiagram} where we observe three (V,V) and  two (V,S) states, i.e.\ three unfragmented and two partially fragmented solutions. It turns out that in the region at lower $\beta$ (higher $1/\beta$) the partially fragmented solutions are coordinated, in so far as both groups of agents have an overall preference for the same market. For high $\beta$ one has the opposite situation, and it is those uncoordinated (V,S) solutions together with an unfragmented (V,V) solution that then merge into a single (S,S) state as $p_\mathcal{B}$ is increased.

We note briefly that the various lines shown in Fig.~\ref{TpBphaseDiagram} were detected by solution tracking, e.g.~by carefully varying $p_\mathcal{B},\beta$ and tracking the number and type of solutions; further details can be found in Appendix~\ref{appendixAlgorithm}. The tracking approach is chosen as it is numerically faster and more reliable than the finite $r$ procedure we used in previous figures, avoiding e.g.\ the numerical noise visible in the two loci in Fig.~\ref{pb055transitions}. It is important to remember that the results only provide information about the existence of steady states, not their stability; the latter can be probed only using actual dynamics as discussed below. Fig.~\ref{TpBphaseDiagram} also relates to fixed market biases so trends with changes in these biases cannot be seen; we have checked, however, that the overall structure of the phase diagram remains intact as long as market biases are symmetric. Quantitative trends were explored in our previous work~\cite{aloric2016}, where we saw that the fragmentation region shrinks as markets become increasingly different.

In summary, the diagram in Fig.~\ref{TpBphaseDiagram} shows that for systems with two symmetric markets and two groups of traders with symmetric buying preferences both fragmented and coordinated (or consolidated) steady states exist across a substantial range of values for the intensity of choice $\beta$. Single market dominance happens when the steady state is either unfragmented or weakly/partially fragmented but coordinated: the majority of trades then happen at a single market. On the other hand, markets can coexist, receiving a roughly even share of trades, when the steady state is strongly fragmentated or weakly/partially fragmented but uncoordinated. In the former case both markets are visited by both groups while in the later case an effective market/group loyalty appears.
In the following sections we analyse these different steady states further, with regards to the average population returns they produce and their stability in simmulated systems with finite $N$ and $r$.

\subsection{Average Population Returns}

\begin{figure*}[t!]
\includegraphics[width=\textwidth]{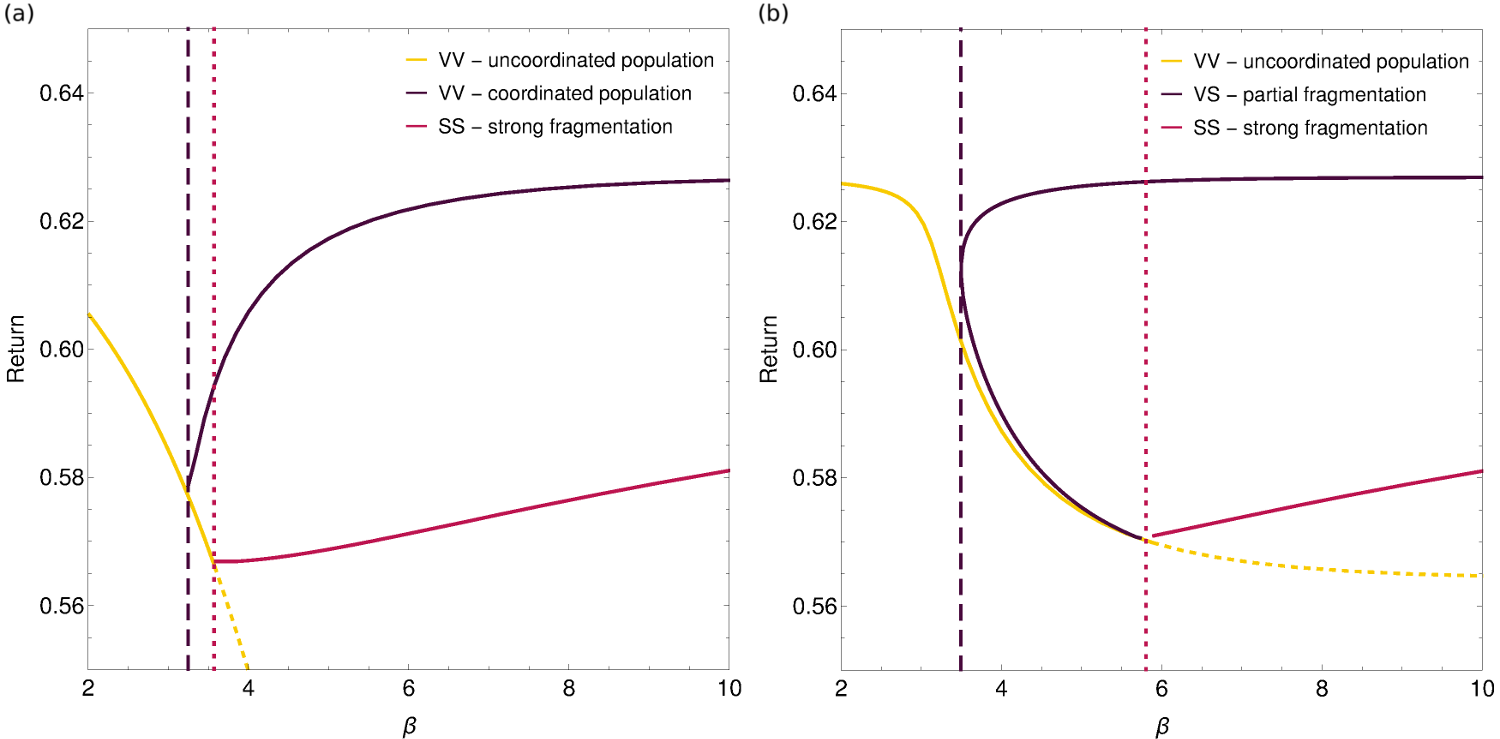}
\caption[Average population returns for different steady states in the $r\rightarrow 0$ limit]{Average population returns for different steady states in the $r\rightarrow 0$ limit. The yellow line shows the low $\beta$ steady state representing uncoordinated population (dashed in the regime where it is no longer a bona fide steady state).
The dashed dark violet line marks the value of $\beta$ where the multiplicity of solutions changes (see Fig.~\ref{TpBphaseDiagram}); at the dashed pink line  strongly fragmented steady states first appear.
(a) Decisive population $(p^{(1)}_\mathcal{B},p^{(2)}_\mathcal{B})=(0.8,0.2)$. The dark violet line shows the average population return for a coordinated unfragmented or weakly fragmented steady state and the pink line similarly for a strongly fragmented state. (b) Indecisive population $(p^{(1)}_\mathcal{B},p^{(2)}_\mathcal{B})=(0.55,0.45)$. The dark violet line gives the average population returns for partially fragmented steady states (coordinated on top, uncoordinated on bottom) and the pink line similarly for a strongly fragmented state.}
\label{r0returns}
\end{figure*}

The phase diagram in Figure~\ref{TpBphaseDiagram} reveals a plethora of possible steady states in the system of two markets and a large population of traders, depending on the traders' learning parameter $\beta$ and their propensity to act as buyers $p_\mathcal{B}$. We now investigate whether these steady states induce differences in average population returns as we saw in small systems, e.g. Figs~\ref{toymodelphasediag} and \ref{4playersPD}. We look at the average population return per trading round, where we count also zero returns that arise from an order being invalid or no trading partner being available.

In Figure~\ref{r0returns}, we show average population returns for the two scenarios of decisive ($(p^{(1)}_\mathcal{B},p^{(2)}_\mathcal{B})=(0.8,0.2)$, panel (a)) and indecisive ($(p^{(1)}_\mathcal{B},p^{(2)}_\mathcal{B})=(0.55,0.45)$, panel (b)) traders. The $\beta$-dependences reflect the transitions between solution types we saw earlier (in Figures~\ref{pb08transitions}, \ref{pb055transitions} and the phase diagram  Fig.~\ref{TpBphaseDiagram}).
The overall trends resemble those for finite $N$. First, we note that the return of the uncoordinated, low $\beta$ solution (marked in yellow in Fig.~\ref{r0returns}) is the lowest among the alternatives once multiple solutions exist. Second, the coordinated states (dark violet) lead to the highest average return. Interestingly, this is not influenced by the type of fragmentation, i.e.\ it is true for both weakly fragmented and partially fragmented states as long as a majority of the population develops a preference for a single market. By comparison, the strongly fragmented state (pink) always leads to a lower average population return.

The differences in the returns achieved by populations of decisive and indecisive traders, respectively, are driven mainly by the fact that indecisive groups can sustain more trades without requiring the presence of other groups at a market. This is particularly visible in the higher population average return for low $\beta$; in this range the decisive population suffers from the group-specific market preferences that tend to separate traders towards different markets and consequently result in a lower number of trades. Additionally, the continuation of the low $\beta$ solution is a viable steady state for a broader range of intensities of choice for the indecisive population. The dashed yellow line marks the region of $\beta$ for which this fixed point is no longer a genuine steady state, as the free energy has multiple minima when evaluated at the order parameters calculated for this fixed point. Along this line the indecisive population return again does not drop as far as it does in the case of a more decisive population. 

In the right panel of Fig.~\ref{r0returns} we note the occurence of the saddle node biffurcation in the transition of the indecisive population, with four new (V,S)-solutions (which come in two pairs giving identical returns) emerging at once. The top branch corresponds to the average population returns at the coordinated partially fragmented states; for greater values of $\beta$ (outside the range shown) these states smoothly transition into weakly fragmented, coordinated, states. The bottom branch relates to uncoordinated partially fragmented states that merge into the strongly fragmented (S,S) state for greater $\beta$. 

Interestingly, the average population return in the high $\beta$ limit of the coordinated state also corresponds to the average population return when all traders choose randomly (i.e. $\beta=0$). This is true because in both limits the average number of agents trading at each market is equal. Intriguingly, this means that when learning is introduced, for low intensities of choice, an agent who takes decisions based on their previous history may be worse off than an agent who plays at random. This effect disappears again only in the large $\beta$ limit of the weakly fragmented state, though
note that in the latter case one group earns more than the other.  
Returning to the strongly fragmented state, despite indications that for a given $\beta$ this is best among the states that do not distinguish between groups in the long run (see Fig 6 of~\cite{aloric2016}), in terms of average population return this state is outperformed by random traders ($\beta=0$).

\subsection{Dynamics}
We now ask what effect the existence of multiple steady states, as predicted by theory for infinite populations, has on the dynamics. We simulate the dynamics numerically, necessarily for finite $N$ and with learning rate $r>0$, i.e.\ for finite memory length $1/r$.
In previous work we have already shown that the theory predicts the steady state properties of finite populations quite well (see e.g.\ Figure 4 in~\cite{aloric2016}).
The role of $r$ is more important as this can shift phase boundaries~\cite{aloric2016}. (Conceptually, the precise distinction for $r\to0$ between weakly and strongly fragmented states is also lost for $r>0$ and becomes a crossover.)

\begin{figure}[h!]
\includegraphics[width=0.45\textwidth]{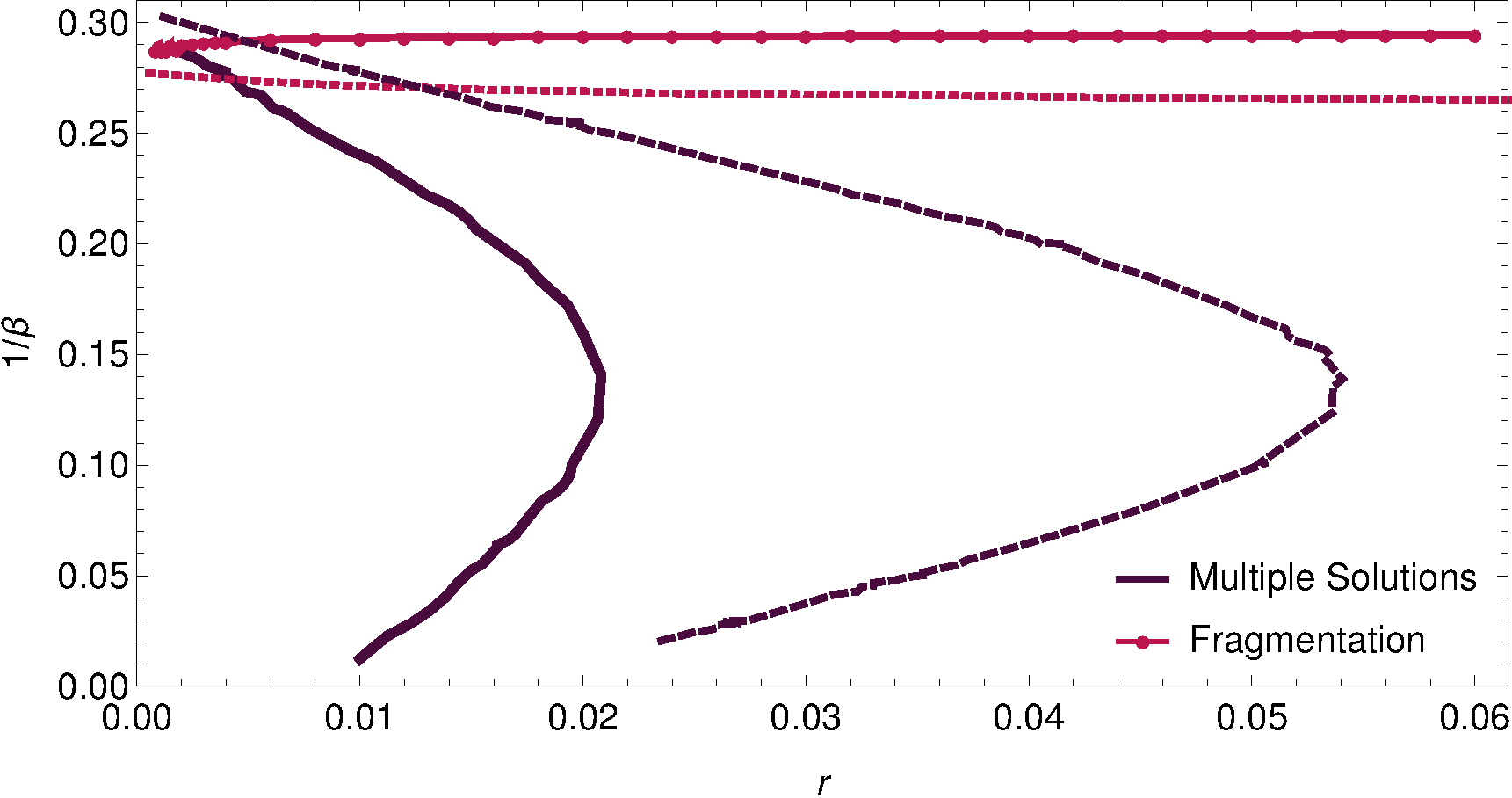}
\caption{Memory length dependence of phase boundaries. The pink line (solid with circles and dotted) shows the fragmentation threshold, where at least one steady state is fragmented (weakly or strongly). The dark violet line (dark solid and dashed) shows the boundary of the region where multiple steady states exist. Solid lines represent indecisive traders $(p^{(1)}_\mathcal{B},p^{(2)}_\mathcal{B})=(0.55,0.45)$ and dashed lines decisive traders $(p^{(1)}_\mathcal{B},p^{(2)}_\mathcal{B})=(0.8,0.2)$. Multiple steady states exist for small enough $r$, i.e.,\ long enough memory $1/r$. The market parameters are $(\theta_1,\theta_\mo)=(0.3,0.7)$.
}
\label{rTphasediag}
\end{figure}
\begin{figure*}[th!]
\includegraphics[width=\textwidth]{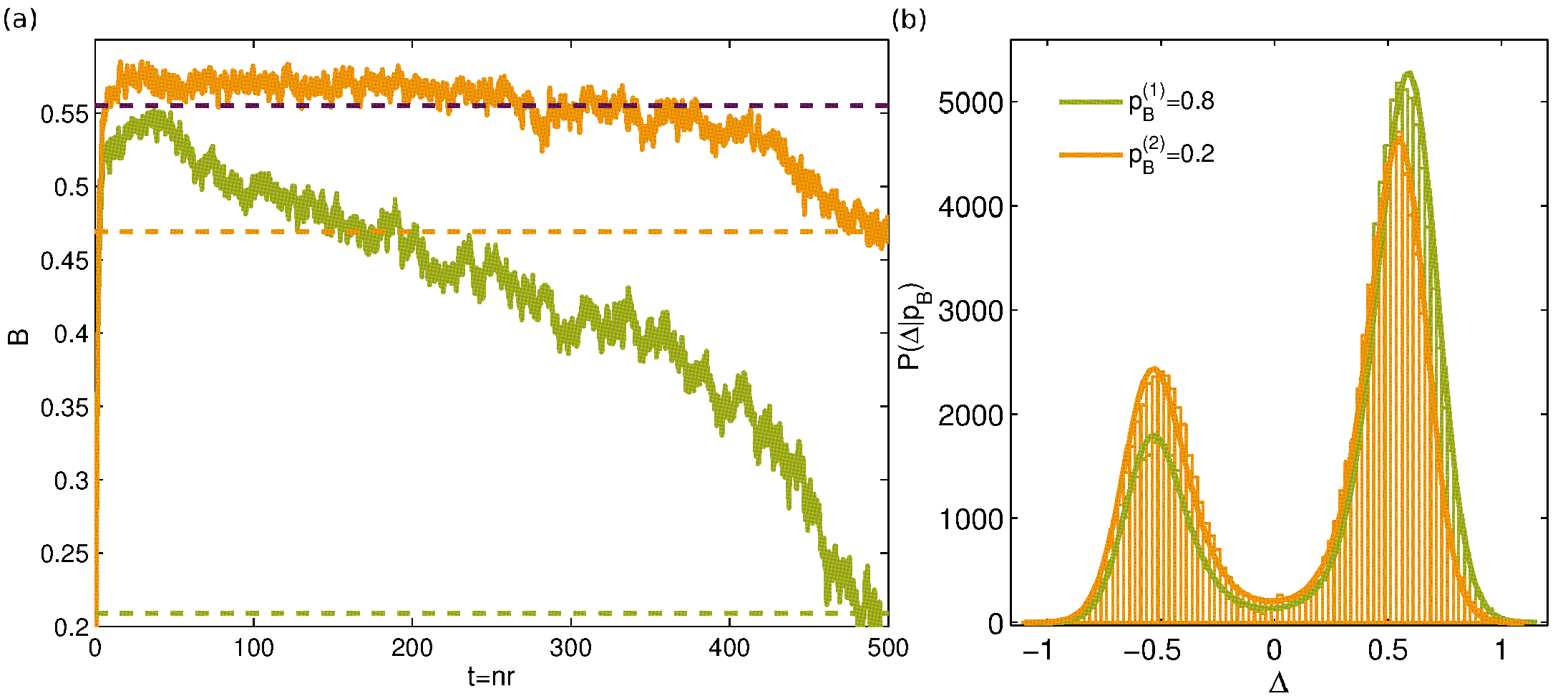}
\caption[Metastability of strongly fragmented state and transition to weakly fragmented state]{Metastability of the strongly fragmented state and transition to the weakly fragmented state: dynamical evolution of a system with $N=2000$ agents using $(r,1/\beta)=(0.05,0.16)$, with preferences for buying $(p_\mathcal{B}^{(1)},p_\mathcal{B}^{(2)})=(0.8,0.2)$, and market parameters $(\theta_1,\theta_2)=(0.3,0.7)$. (a) Evolution of Binder cumulants of the two attraction distributions of the two agent groups [buyers (green) and sellers (orange)]. Dashed lines are theoretical predictions for the strongly fragmented steady state (dark violet, equal for both groups) and weakly fragmented state (green and orange for the two groups). (b) Attraction distributions predicted from theory for the weakly fragmented steady state (solid line) compared to simulation data at $t=500$ (histogram).}
\label{BinderSimulation}
\end{figure*}

In Figure~\ref{rTphasediag} we illustrate the $r$-dependence of two key phase boundaries, for the two populations we have mainly considered so far (decisive $p_{\mathcal B}=0.8$ and indecisive $p_{\mathcal B}=0.55$). We note that the region of multiple steady states shrinks with increasing $r$ for both populations while the fragmentation line is only weakly $r$-dependent. The lines are related to the lines of the same color in the $(\beta,p_\mathcal{B})$ phase diagram in Fig.~\ref{TpBphaseDiagram}
and the dashed grey lines marked in Fig.~\ref{TpBphaseDiagram} correspond to the $r\to 0$ limit of the $(r,\beta)$ phase diagram in Fig.~\ref{rTphasediag}.

Overall, Figure ~\ref{rTphasediag} tells us that we need to use reasonably small $r$, certainly below $0.05$ for $p_{\mathcal B}=0.8$, to see multiple steady states in numerical simulations. As smaller $r$ slow the dynamics, we choose in practice values of $r$ that are as large as possible while staying well within the multiple states regime.

In Figure~\ref{BinderSimulation} we show numerical data for the actual dynamics of a system of decisive tradners $(p^{(1)}_\mathcal{B},p^{(2)}_\mathcal{B})=(0.8,0.2)$ at our standard market parameters $(\theta_1,\theta_2)=(0.3,0.7)$, taken from a single run for a population with $N=2000$ traders (see~\cite{aloric2017thesis} for simulation details) using learning rate and inverse decision strength $(r,1/\beta)=(0.05,0.16)$.
For these parameters the phase diagram of Fig.~\ref{rTphasediag} predicts the existence of three steady states, two weakly fragmented states (with the majority of both groups coordinated at the same market, $m=\mo$ or $m=1$) and a strongly fragmented state (this state was studied in~\cite{aloric2016}, see Fig.\ 3 there; it is the unique steady state for the larger $r=0.1$ used in~\cite{aloric2016}). As a global summary statistic of the shape of the attraction distributions of the two groups of agents we use the Binder cumulant \cite{Binder1981}
\[
B=1-\frac{\langle \Delta^4\rangle_{P(\Delta)}}{3\langle \Delta^2\rangle^2_{P(\Delta)}}
\]
and plot this over time (see further discussion in \cite{aloric2016,aloric2017thesis}). Away from the strongly fragmented state the attraction distributions of the two groups are not related by a symmetry so we plot their Binder cumulants separately.

Figure~\ref{BinderSimulation} shows that the system quickly reaches the strongly fragmented state, with the Binder cumulants being close to the theoretically predicted value; the slight deviation can be attributed to the finite population size.
The dynamics then branches off from the theoretical prediction at $t\approx 50$, showing that the strongly fragmented state is, for finite $N$, only metastable. The departure is led by one of the agent groups and reaches one of the theoretically expected weakly fragmented states at $t\approx 500$, as shown in Figure~\ref{BinderSimulation} by the agreement both of the relevant Binder cumulants and the full attraction distributions (b). 

\begin{figure*}[ht!]
\includegraphics[width=\textwidth]{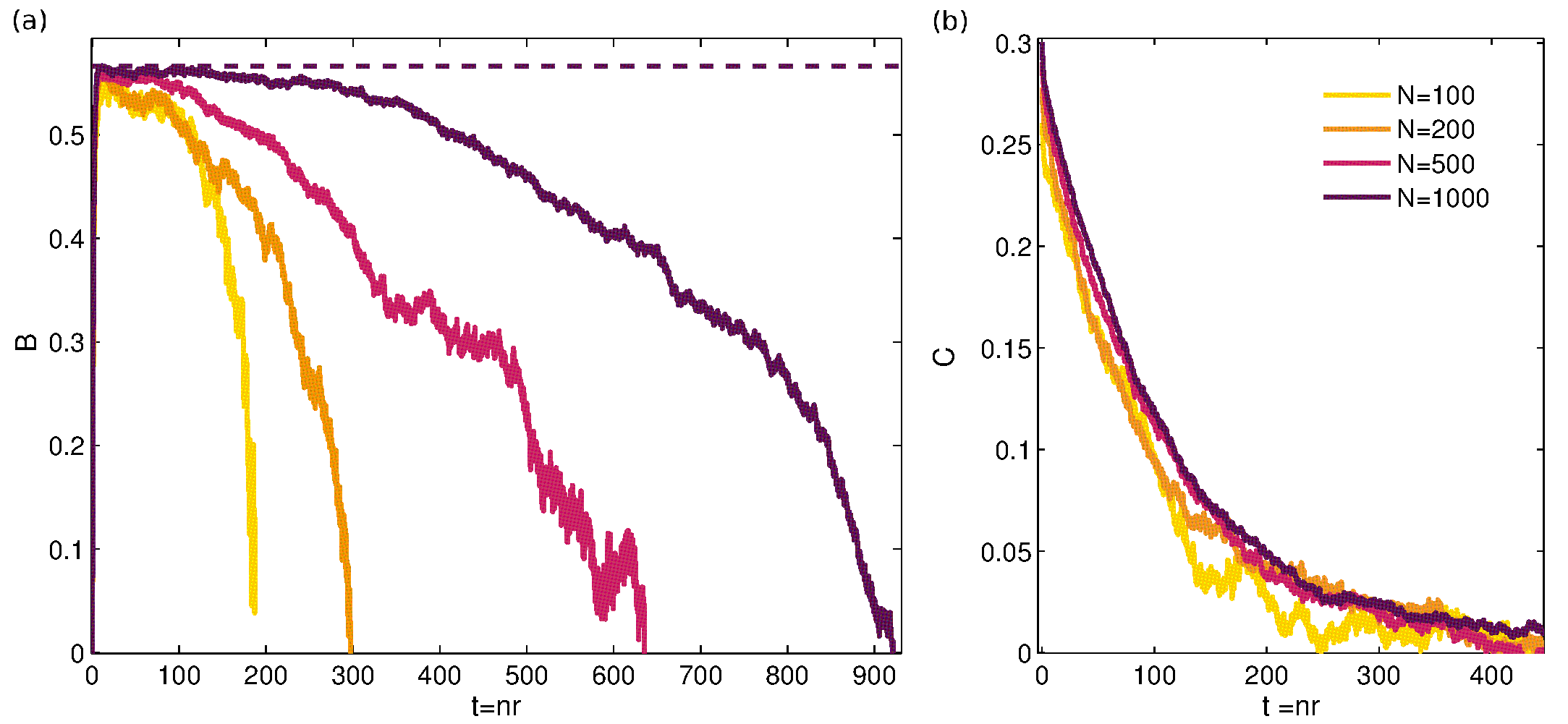}
\caption[Lifetime of strongly fragmented state for different system sizes $N$]{Lifetime of strongly fragmented state for different system sizes $N$. (a) Binder cumulant time series (averaged over the two agent groups for compactness) along with the $N\rightarrow\infty$ theoretical prediction for the strongly fragmented state (dashed line), showing an increase of lifetime with $N$. $1/\beta=0.15$ while the other parameters are as in Fig.~\ref{BinderSimulation}. (b) Autocorrelation function of single agent attraction differences $C(t)=\langle(\Delta^i(\tau)-\overline{\Delta(\tau)})(\Delta^i(\tau+t)-\overline{\Delta(\tau+t)})\rangle$: the single agent autocorrelation time is essentially $N$ independent.
}
\label{sslifetimeN}
\end{figure*}

We proceed in Fig.~\ref{sslifetimeN} (a) to analyse the life time of the strongly fragmented steady state in more detail.
The figure displays Binder cumulant time series for different population sizes at the same learning parameters $(r,1/\beta)=(0.05,0.15)$ and shows that the lifetime increases with system size (we have not analysed the $N$-dependence in detail; in the range shown it is approximately linear). We can compare this with the time correlations of the attraction difference $\Delta$ for individual agents: Fig~\ref{sslifetimeN} (b) graphs this correlation function, measured from the point in time when the strongly fragmented state is first reached. One sees clearly that the single agent correlation time is essentially independent of $N$ while the lifetime of the strongly fragmented state grows significantly with system size $N$. The conclusion is that strong fragmentation is a long-lived state of the population for large $N$, within which single agents effectively ``equilibrate'' by losing all memory of their initial preferences.

\begin{figure}[h]
\includegraphics[width=0.48\textwidth]{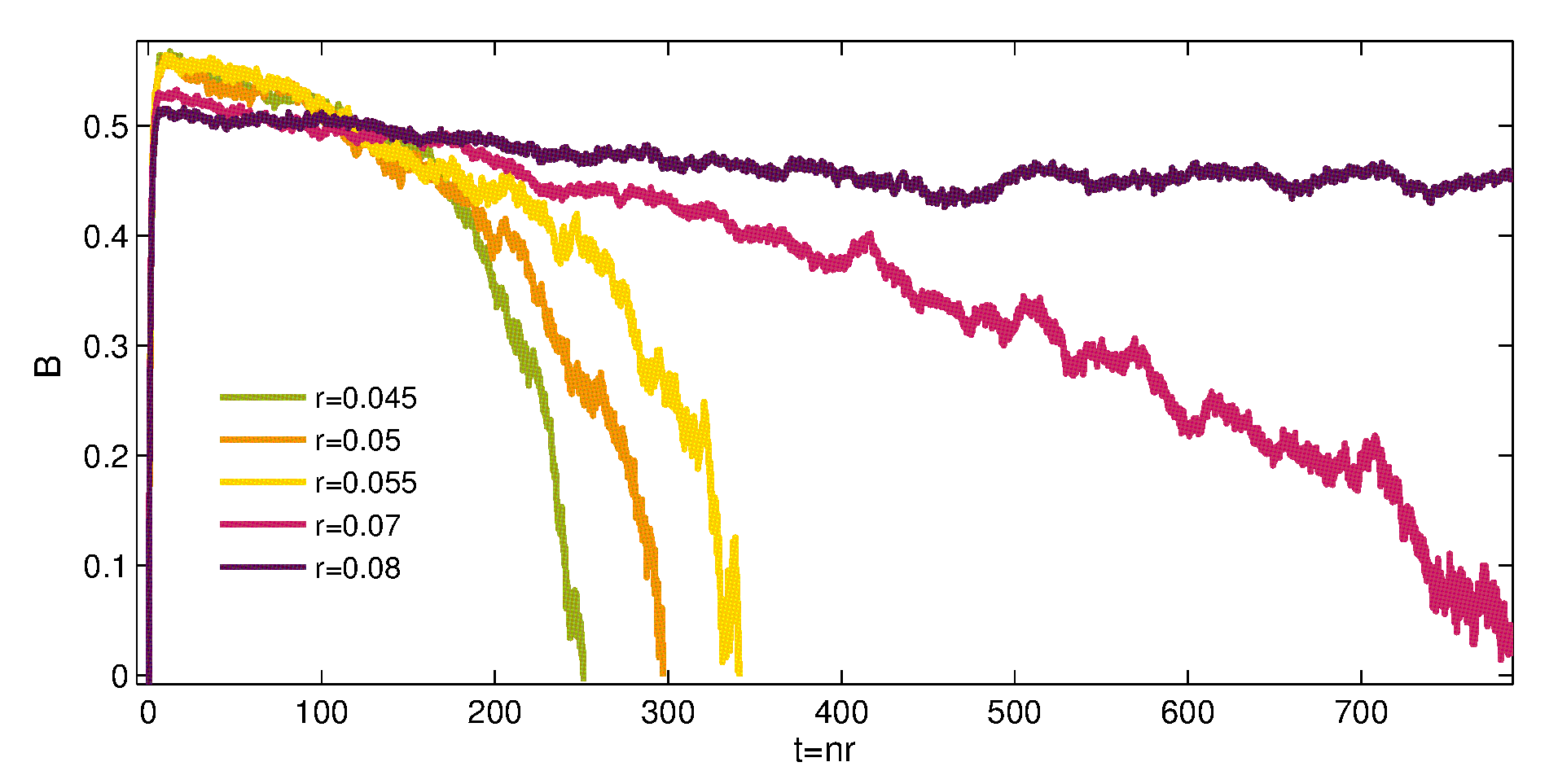}
\caption[Binder cumulant time series for different learning rates $r$ at the fixed intensity of choice]{Binder cumulant time series for different learning rates $r$ at the fixed intensity of choice $1/\beta=0.15$. All parameters are as in Fig.~\ref{BinderSimulation}, except for the smaller population size $N=200$ and $r$ as shown. The lifetime of the strongly fragmented state lifetime increases with $r$, eventually becoming infinite when this state is the only steady-state solution.
}
\label{sslifetimeR}
\end{figure}

In Figure~\ref{sslifetimeR} we move to the $r$-dependence of the lifetime of the strongly fragmented state, showing Binder cumulants for a small system $N=200$ for different $r$-values at fixed $1/\beta=0.15$. 
For all values of $r$, rapid initial convergence to the strongly fragmented state is observed. Within this state the Binder cumulants depend weakly on $r$ as has been noted previously~\cite{aloric2016}), reflecting the $r$-dependence of the attraction distributions.
The lifetime of the strongly fragmented state, set by the decay of the Binder cumulant to lower values, {\em increases} with $r$. This is consistent with the results of Fig.~\ref{rTphasediag}, which showed that above some $\beta$-dependent threshold value for $r$ the strongly fragmented state is the only steady state and thus must be stable, corresponding to an infinite lifetime. For the value $\beta=1/0.15$ in Fig.~\ref{sslifetimeR}, theory predicts this threshold to be $r\approx 0.055$. Numerically we see that the strongly fragmented state has a finite lifetime up to $r=0.07$, presumably due to finite population effects for the relatively small $N=200$ used in the figure.

We find qualitatively the same features as above also in numerical simulations of the dynamics of a system of indecisive traders, with populations first reaching a long-lived (for large $N$) strongly fragmented state and eventually decaying into a partially fragmented state. This is the behaviour when such multiple steady states are predicted by our theory, i.e.\ for small $r$; for larger $r$ (above $r_c\approx 0.02$, see Fig.~\ref{rTphasediag}) strong fragmentation is the only steady state. Quantitatively, we find that where strong fragmentation is metastable its lifetimes are significantly longer than for the decisive traders, exceeding our maximal simulation times of $10^6$ trading rounds ($t=20000$) for the largest $r<r_c$.

We comment finally on the role of initial conditions. In the dynamical simulations shown so far we 
used for these $P(\Delta|p_\mathcal{B})=\delta(\Delta)$, corresponding to the reasonable assumption that the agents have no initial preference for either market. We also explored Gaussian initial distributions for the attraction differences, $P(\Delta|p_\mathcal{B})=\mathcal{N}(\mu,\sigma^2)$. Where there is only a single steady state we then find, as expected, that this state is reached irrespective of the chosen initial condition. On the other hand, where the theory predicts multiple steady states, the initial conditions do matter. We observe that the metastable strongly fragmented state continues to be reached whenever the mean initial attraction difference $|\mu|$ is small enough, irrespective of the standard deviation $\sigma$. As $|\mu|$ is increased we see that the dynamics ``misses'' the metastable strongly fragmented state and rapidly moves to a final weakly or partially fragmented state. This is consistent with the intuition that these states break the symmetry between markets, and hence are favoured when the population already starts off with an overall initial preference for one of the markets.

\section{Discussion \& Conclusions}
In this paper, our aim was to investigate the existence of coordinated and fragmented steady states in a system of agents choosing adaptively between two markets. We focussed primarily on the long memory limit, where the transition to fragmentation is sharp. We first studied two traders who learn how to coordinate at a market and maximise their average return even though one of them will necessarily earn less. Moving to a four player system we observed fragmentation in addition to coordination. Interestingly, we found that coordinated and fragmented states lead to the same average population return for high intensity of choice $\beta$, in spite of the presence of two different types of agents (buyers and sellers). In the coordinated state one of the agent types will always earn less, while in the fragmented state both types have the same average, but one agent from each group is less satisfied. Thus at the fragmented state, average returns do not discriminate between types of agents.

We then introduced a general method for determining the type and number of steady states in the limit of large populations with long memory. This can be done in our setup with only a single order parameter per market. After a preliminary analysis for exogenously determined order parameters, we saw that in the general case a self-consistency criterion determines the order parameters in the steady state. Analysing a quantity analogous to a free energy then allows one to say whether a population (or one of its groups) is fragmented and whether this fragmentation is strong or weak.

Already for small system sizes we noticed that the agents' preference for buying, $p_\mathcal{B}$, is an important system parameter. Not only does it influence the critical intensity of choice $\beta$ on $p_\mathcal{B}$ for the onset of fragmentation, but for $N\geq 4$ it also qualitatively affects the nature of the steady states. This remains true also for the $N\rightarrow \infty$ limit, where we find a rich variety of steady states in the $(\beta,p_\mathcal{B})$ diagram, in spite of the simplified nature of our models for markets and traders. These include: market coexistence - where both markets attract both types of traders (S,S), and where market/trader specialisation occurred (W,W) (uncoordinated weakly fragmented state for moderately indecisive traders); single market dominance (W,W) (coordinated weakly fragmented states); market indifference (U,U) (e.g.\ for low $\beta$); general vs.\ specialised markets (e.g.\ (U,S), where a single market attracts both groups of agents while the other can be viewed as specializing towards only one group). Interestingly, all these different steady states arise without imposing any heterogeneity onto the agents (in contrast to assumptions elsewhere~\cite{gomber2017competition}) and fragmentation is the preferred state even when the markets have identical properties (contrary to views expressed in~\cite{pagano1989trading,Chowdhry1991}).

To interpret our results for the prevalence of fragmentation more broadly we can draw on the work of Cheung et al.~\cite{cheung1997}, who use evidence from Behavioural Game Theory to suggest that values of $\beta$ are consistent across games but increase in more informative environments. The authors also argue that a parameter closely analogous to $r$ increases with the trustworthiness of information in the system. Bearing in mind the results shown in Fig.~\ref{rTphasediag}, where for large $r$ and large $\beta$ the only steady state is the fragmented one, this suggests that more informative environments, or ones where information is more trustworthy because e.g.\ of stability over long timescales, might naturally
lead to fragmented states. The prevalence of the strongly fragmented state is clear also from Figure~\ref{TpBphaseDiagram}, which shows that this state exists for all populations with groups symmetrically biased towards buying and selling, respectively.

One of the non-trivial predictions of our theory is the existence of partiallay fragmented states, where one group of agents (e.g.\ those who have a preference for buying) fragments while the other (where agents prefer to sell) does not. We saw that the region in the phase diagram where such states appear increases with $N$ for indecisive traders and shrinks for decisive traders (compare Fig.~\ref{4playersPBdiag} for $N=4$ and Fig.~\ref{TpBphaseDiagram} for $N\to\infty$).

We studied also the average population returns achieved by agents in the various steady states. For large populations we saw that the coordinated weakly fragmented steady state leads to the highest population average returns, even though one agent group earns less in that state. We also noticed that such steady states, which essentially represent coordination at a single market when $r\rightarrow 0$, lead to the same average payoff for large $\beta$ as for random agents ($\beta=0$). This is because coordination at a single market, just like random market choice, leads to the same number of buyers and sellers at a single market and thus the same number of successful trades and average returns. Interestingly, this shows that weak learning (finite $\beta$) leads to lower returns, e.g.\ not choosing the strictly best trading venue (in terms of returns) can be worse for an agent than random guessing. This behaviour is rather similar to the ``J-curve'' effect studied in~\cite{huber2007,toth2007b} where, in the context of trading agents with different information levels, moderately informed agents earn less from higher informed agents but also from uninformed, randomly trading, agents.

Finally we investigated, by means of numerical simulations, how the theoretically predicted steady states appear in the dynamics of finite agent populations. 
If the agent start as ``blank canvasses'' (without initial market preference), we found that the adaptation process always leads to the strongly fragmented state first. This state is metastable, with a lifetime that grows large with population size, and the system eventually settles into one of the weakly fragmented states. This remains true even if there is scatter in the agents' initial preferences, while a systematic initial bias towards one of the markets can cause the dynamics to ``miss out'' the metastable strongly fragmented state.
To put this result into more intuitive terms, two markets that enter into competition to attract on average indifferent traders will always exhibit a period of coexistence in a strongly fragmented state (and if $r>r_c$ this coexistence will last indefinitely), whereas if the population is not indifferent initially then a market monopoly will arise much more quickly.

\paragraph*{Acknowledgements.} The authors are grateful to Peter McBurney and Robin Nicole for useful discussions. PS acknowledges the stimulating research environment provided by the EPSRC Centre for Doctoral Training in Cross-Disciplinary Approaches to Non-Equilibrium Systems (CANES, EP/L015854/1). AA acknowledges the support of the Ministry of Education, Science, and Technological Development of the Republic of Serbia under Project ON171017.
\bibliography{saskabib}
\appendix

\section{\label{FPappendix}Details of the Fokker-Planck description}

In this appendix we provide some of the explicit expressions appearing in the Fokker-Planck description of our market choice model. As per definitions of bid and ask distributions and score assignments, defined in \textit{Trading strategies} of section~\ref{sec:model}, the return distributions for an agent choosing a market $m$ and an order type $\mathcal{B}$ or $\mathcal{S}$ are:
\begin{widetext}
\begin{align}
P(S|m,\mathcal{B})
& =
Q_{\mathcal{B}m}T_{\mathcal{B}m}
\frac{1}{Q_{\mathcal{B}m}\sigma_b\sqrt{2\pi}}
\exp\left(-\frac{(S-(\mu_b-\pi_m))^2}{2\sigma_b^2}\right)
\theta(S)
+\delta(S)(1-Q_{\mathcal{B}m}T_{\mathcal{B}m}),\nonumber\\
P(S|m,\mathcal{S})& =
\underbrace{Q_{\mathcal{S}m}T_{\mathcal{S}m}}_\text{agent trades}\underbrace{\frac{1}{Q_{\mathcal{S}m}\sigma_a\sqrt{2\pi}}
\exp\left(-\frac{(S-(\pi_m-\mu_a))^2}{2\sigma_a^2}\right)
\theta(S)}_\text{non-negative return}
+\delta(S)\underbrace{(1-Q_{\mathcal{S}m}T_{\mathcal{S}m})}_\text{agent does not trade}.
\end{align}
\end{widetext}
(Note that in statements of these distributions in previous publications~\cite{aloric2016}, $\mu_a$ and $\mu_b$ were omitted due a typographical error.) 
When agents have fixed buying preferences $p_{\mathcal{B}}$, their return distribution is then dependent only on the chosen market $m$:
\begin{align*}
P(S|m)=p_\mathcal{B}P(S|m,\mathcal{B})+(1-p_\mathcal{B})P(S|m,\mathcal{S}).
\end{align*}
The probabilities that an order is valid, $Q_\gamma$, are given by
\begin{align*}
Q_{\mathcal{B}m}&=\frac{1}{\sigma_b\sqrt{2\pi}}\int_{\pi_m}^{\infty}db \,\exp\left(-\frac{(b-\mu_b)^2}{2\sigma_b^2}\right),\nonumber\\
Q_{\mathcal{S}m}&=\frac{1}{\sigma_a\sqrt{2\pi}}\int_{-\infty}^{\pi_m}da \,\exp\left(-\frac{(a-\mu_a)^2}{2\sigma_a^2}\right).
\end{align*}
and can be expressed in terms of error functions~\cite{aloric2017thesis}. 

\begin{figure*}[ht!]
\includegraphics[width=0.9\textwidth]{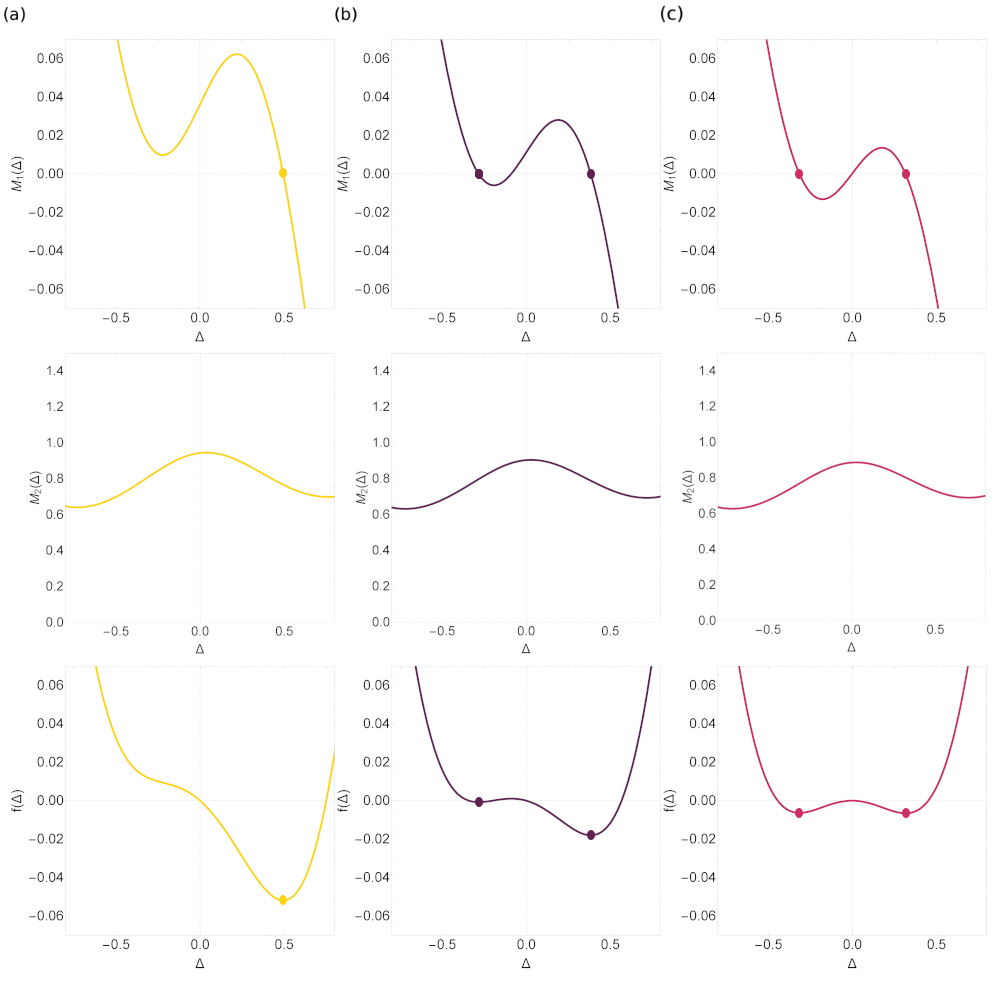}
\caption{Drift $M_1(\Delta)$ (top), diffusion $M_2(\Delta)$ (middle) and free energy $f(\Delta)$ (bottom) functions for a subgroup with preference for buying $p_\mathcal{B}=0.8$. Function plots illustrate three qualitatively different conditions for the following pairs of market order parameters: (a) $(D_1,D_\mo)=(1,1)$, unfragmented region; (b) $(D_1,D_\mo)=(1.1,1)$, weakly fragmented region; and (c) $(D_1,D_\mo)=(1.15,1)$, strongly fragmented region. Market biases are set to the standard values $(\theta_1,\theta_\mo)=(0.7,0.3)$. All functions are evaluated at the intensity of choice $\beta=1/0.265$. }
\label{moments}
\end{figure*}
The transition kernel between two states $\Delta$ and $\Delta'$ of an agent with buying preference $p_\mathcal{B}$ is
\begin{widetext}
\begin{align}
K(\Delta'|\Delta,p_\mathcal{B})= \int dS 
\sum_{m=\mo}^1
\Big[p_\mathcal{B}P(S|m,\mathcal{B})+(1-p_\mathcal{B})P(S|m,\mathcal{S})\Big]P(m|\Delta)\delta(\Delta'-mrS-(1-r)\Delta).
\end{align}
\end{widetext}
The resulting drift and diffusion terms for small $r$ are discussed in detail in~\cite{aloric2017thesis}, here (Fig.~\ref{moments}) we provide plots corresponding to Eqs.~(\ref{firstmoment},\ref{secondmoment}), evaluated at three different sets of market order parameters for ilustration. We consider the value  $\beta=1/0.265$ for the intensity of choice, in order to match Fig.~\ref{pb08transitions} (c). The three sets of market order parameters all lie on a horizontal line ($D_\mo=-1$), while $D_1$ is changed so that the order parameters lie in the unfragmented, the weakly fragmented or the strongly fragmented region respectively. Plots in the left panels (a) ilustrate market conditions leading to an unfragmented distribution -- there is a unique solution of $M_1(\Delta|p_\mathcal{B},T_\gamma)=0$, corresponding to the unique free energy minimum (calculated from Eq.~\ref{freeenergy} and shown in the third row of the figure). Both are marked by a circle. The middle panels (b) ilustrate the weakly fragmented case, where there are three zeros of the drift term (two stable fixed points and one unstable one), corresponding to two minima of the free energy; as the minima are at different height the resulting (steady state) distribution of $\Delta$ will become concentrated around the lowest minimum for $r\rightarrow 0$ as discussed in the main text. Finally, the case shown in the panels (c) has two equal minima of the free energy and thus represents a strongly fragmented scenario. Note that the diffusion term $M_2(\Delta)$ is in all three cases of order unity and does not affect the number of free energy minima; it only makes a quantitative contribution to the free energy and hence to $P(\Delta|p_\mathcal{B},T_\gamma)$.

\section{\label{appendixAlgorithm}Algorithmic remarks}
The method of finding all steady state solutions by identifying loci of self-consistent market order parameters is the best way to exhaust market order parameter space and thus to find all the solutions for the finite $r$. By identifying the domains where these solutions lie we can also fully characterize the solution at nonzero $r$, obtaining information about the limit $r\to 0$ by extrapolation. However, this method is numerically demanding as for every point in order parameter space we need to find a steady state distribution (its normalisation usually takes most of the processing time) and recalculate the corresponding order parameters. Checking what corrections arise for $r\to 0$ takes additional time.
We describe numerically less demanding alternatives below.
\paragraph*{Population with homogeneous market preferences.} We have seen that depending on system parameters the attraction distribution for a group of agents can be unimodal in the $r\rightarrow 0$ limit (``U'' and ``W'' states). These states represent a population where the market preferences within the group are homogeneous. This realisation offers a straightforward way to find all the states of this type for any system parameter. The demand-to-supply order parameters simplify to
\begin{widetext}
\begin{align}
D_m&=\frac{p_\mathcal{B}^{(1)}\int d\Delta \sigma_\beta(m\Delta) P(\Delta|p_\mathcal{B}^{(1)})+p_\mathcal{B}^{(2)}\int d\Delta \sigma_\beta(m\Delta) P(\Delta|p_\mathcal{B}^{(2)})}{(1-p_\mathcal{B}^{(1)})\int d\Delta \sigma_\beta(m\Delta) P(\Delta|p_\mathcal{B}^{(1)})+(1-p_\mathcal{B}^{(2)})\int d\Delta \sigma_\beta(m\Delta) P(\Delta|p_\mathcal{B}^{(2)})}\nonumber\\
&=\frac{p_\mathcal{B}^{(1)}\sigma_\beta(m\Delta^{(1)})+p_\mathcal{B}^{(2)}\sigma_\beta(m\Delta^{(2)})}{(1-p_\mathcal{B}^{(1)})\sigma_\beta(m\Delta^{(1)})+(1-p_\mathcal{B}^{(2)})\sigma_\beta(m\Delta^{(2)})}\ ,
\label{hompopulationOP}
\end{align}
\end{widetext}
where in the second row we have used $\langle\sigma_\beta(\Delta)\rangle=\sigma_\beta(\langle\Delta\rangle)$, a relation that is exact in the $r\rightarrow 0$ limit where the steady state distribution is a delta distribution centred at $\Delta^{(g)}$. To identify these peak positions we find the zeros of the first jump moment $M_1$ as defined in Eq.~(\ref{firstmoment}), taking into account the dependence of $D_m$ on the attraction difference $\Delta^{(g)}$ in each group. This means that when searching for a steady state in which both groups of traders have homogeneous market preferences, we need to solve the peak position equations for the two groups simultaneously:
\begin{align}
M_1^{(1)}(\Delta^{(1)}|p_\mathcal{B}^{(1)},D_m(\Delta^{(1)},\Delta^{(2)}))=0\ ,\nonumber\\
M_1^{(2)}(\Delta^{(2)}|p_\mathcal{B}^{(2)},D_m(\Delta^{(1)},\Delta^{(2)}))=0\ .
\label{populationEq}
\end{align}
Every solution $(\Delta^{(1)*},\Delta^{(2)*})$ found in this way needs to be checked for consistency with the initial assumption of homogeneous market preferences, \i.e. the market order parameters corresponding to every solution pair need to belong to the unfragmented or weakly fragmented solution domain. This is done by  calculating the corresonding order parameters $D_m$ from Eq.~(\ref{hompopulationOP}) and finding the ``free energy'' corresponding to these order parameters. If the global free energy minimum is centred at $\Delta^{(g)*}$ the solution is consistent with our initial assumption and we have found a (homogeneous) population steady state. Depending on the signs of $\Delta^*$ we classify such steady states further as either \textit{coordinated} for $\Delta^{(1)*}\Delta^{(2)*}>0$ or \textit{uncoordinated} for $\Delta^{(1)*}\Delta^{(2)*}<0$. For any finite intensity of choice $\beta$, a single agent can of course choose another market even if the state is categorized as coordinated at market 1, but the categorization is exact for the $\beta\rightarrow \infty$ limit.

In the second case study ($p_\mathcal{B}=0.55$), the continuation of the low $\beta$ fixed point is a solution we can consistently find by this method for a wide range of intensities of choice, much wider than when the groups have more pronounced buy/sell preferences. 
Crossing the dark violet line in phase diagram (Fig.~\ref{TpBphaseDiagram}), the new fixed points that arise turn out to be all inconsistent with the homogeneous population assumption until very high intensities of choice. This is why we need to employ different techniques to find the other solutions presented in Figure~\ref{pb055transitions}. Only when the intensity of choice is increased further do partially fragmented states cease to exist, and solutions consistent with the homogeneous population assumption return. 

\paragraph*{Strongly co-fragmented state (S,S).} 
To find if these states exist we apply a procedure based on a Maxwell construction argument outlined in Section~\ref{sec:singlepopulation} and in~\cite{aloric2017thesis} for a population consisting of a single group. For each group we define a locus in the space of order parameters $(D_1,D_2)$ for which the strong fragmentation condition Eq.\~(\ref{segregationCondition}) is satisfied. If there is an intersection $(D_1^*,D_2^*)$ between the two loci there are market demand-to-supply ratios in which both groups favour a strongly fragmented state. We finally need to confirm that the two order parameters can be created if only the two fragmented groups trade on the markets. If we assume the strongly fragmented distributions are of the form
\begin{align*}
P(\Delta|p_\mathcal{B}^{(g)})=\omega^{(g)}\delta(\Delta-\Delta_1^{(g)})+(1-\omega^{(g)})\delta(\Delta-\Delta_2^{(g)})
\end{align*}
then the corresponding order parameters are:
\begin{widetext}
\begin{align}
&D_m =\frac{N_{\mathcal{B}m}}{N_{\mathcal{S}m}}\nonumber\\
&D_m =\nonumber\\
&\frac{p_\mathcal{B}^{(1)}\left[\omega^{(1)}\sigma_\beta(m\Delta_1^{(1)})+(1-\omega^{(1)})\sigma_\beta(m\Delta_2^{(1)})\right]+p_\mathcal{B}^{(2)}\left[\omega^{(2)}\sigma_\beta(m\Delta_1^{(2)})+(1-\omega^{(2)})\sigma_\beta(m\Delta_2^{(2)})\right]}
{(1-p_\mathcal{B}^{(1)})\left[\omega^{(1)}\sigma_\beta(m\Delta_1^{(1)})+(1-\omega^{(1)})\sigma_\beta(m\Delta_2^{(1)})\right]+(1-p_\mathcal{B}^{(2)})\left[\omega^{(2)}\sigma_\beta(m\Delta_1^{(2)})+(1-\omega^{(2)})\sigma_\beta(m\Delta_2^{(2)})\right]}
\label{DmFromWeights}
\end{align}
\end{widetext}
If there are weights $\omega^{(g)}\in[0,1]$ corresponding to the intersection point $(D_1^*,D_2^*)$  then the strongly fragmented state exists. These states leave both markets equally active and as we showed in the discussion in Sec~\ref{sec:2populations} they entail benefits for the population as a whole, not favouring any of the symmetric groups.

\paragraph*{Partially fragmented states.} Finally, we outline a procedure to find a population steady state that is a combination of a bimodal (S) state in one group and a unimodal (U or W) state in the other, for $r\rightarrow 0$. A starting point for this search can be obtained by solving the homogeneous population equations~(\ref{populationEq}). When one of the groups is consistent with the homogeneous population assumption while the other is not, we can investigate whether the strongly fragmented solution for this other population exists. To find these states, we assume that the group that is inconsistent with a given homogeneous population solution is in the fragmented state. Thus possible order parameters for this state are on the locus defined by the Maxwell construction. For every pair $(D_1,D_2)$ from the fragmented state locus we investigate the ``free energy'' of the second group (whether it is unfragmented or weakly fragmented). We find the peak position and represent the attraction distribution as a unimodal distribution centred at the (global) free energy minimum. We only need to examine whether by peak weight redistribution of the strongly fragmented group we can retrieve the initial order parameters $(D_1,D_2)$. When this is possible, the partially fragmented state exists. In the example shown in Figure~\ref{pb055transitions}, due to mild buy/sell preferences, when one of the groups is fragmented there are two unfragmented options for the second group, corresponding to specialisation to either of the two markets.

\end{document}